\def\isReadyToSubmit{1}   %

\documentclass[sigplan,10pt]{acmart}

\acmYear{2025}\copyrightyear{2025}
\setcopyright{rightsretained}
\acmConference[EuroSys '25]{Twentieth European Conference on Computer Systems}{March 30--April 3, 2025}{Rotterdam, Netherlands}
\acmBooktitle{Twentieth European Conference on Computer Systems (EuroSys '25), March 30--April 3, 2025, Rotterdam, Netherlands}
\acmDOI{10.1145/3689031.3696096}
\acmISBN{979-8-4007-1196-1/25/03}

\usepackage{times}
\usepackage{graphicx}
\usepackage{subfigure}
\usepackage{array} %
\usepackage{amsmath,amsfonts,amsthm}
\usepackage{color}
\usepackage{listings} %
\usepackage{xspace} %
\usepackage[noend]{algpseudocode} %
\usepackage{algorithm} %
\usepackage{wrapfig} %
\usepackage{flushend} %
\usepackage{multirow} %
\usepackage{booktabs} %
\usepackage{anyfontsize} %
\usepackage{paralist} %
\usepackage[font=small]{caption}
\usepackage{courier} %
\usepackage{verbatim}
\usepackage{calc}
\usepackage{amsthm}
\usepackage{pbox}
\usepackage{algorithm, algpseudocode}

\usepackage[compact,small]{titlesec}

\usepackage{style/macros}

\titlespacing{\section}{0pt}{*.6}{*.2}
\titlespacing{\subsection}{0pt}{*.4}{*.2}
\titlespacing{\subsubsection}{0pt}{*.4}{*.2}

\usepackage{etoolbox}
\makeatletter
\patchcmd{\ttlh@hang}{\parindent\z@}{\parindent\z@\leavevmode}{}{}
\patchcmd{\ttlh@hang}{\noindent}{}{}{}
\makeatother

\let\oldenumerate\enumerate
\renewcommand{\enumerate}{
    \oldenumerate
    \setlength{\itemsep}{.5pt}
    \setlength{\parskip}{0pt}
    \setlength{\parsep}{0pt}
}

\let\olditemize\itemize
\renewcommand{\itemize}{
    \olditemize
    \setlength{\itemsep}{1pt}
    \setlength{\parskip}{0pt}
    \setlength{\parsep}{0pt}
}

\theoremstyle{definition}

\makeatletter
\renewcommand{\ALG@beginalgorithmic}{\footnotesize}
\makeatother

\newcommand{\ie}{{\em i.e.,~}}
\newcommand{\eg}{{\em e.g.,~}}

\def\F{Fig.~}
\def\T{Tab.~}

\newcommand{\ar}[3]{} %
\include{latexdefs}
\include{mathnotation}

\newcommand{\heading}[1]{{\vspace{0pt}\noindent\bf{#1}}} %

{\makeatletter
\gdef\xxxmark{%
    \expandafter\ifx\csname @mpargs\endcsname\relax %
        \expandafter\ifx\csname @captype\endcsname\relax %
            \marginpar{\textcolor{red}{xxx~}}%
        \else
            \textcolor{red}{xxx~}%
        \fi
    \else
        \textcolor{red}{xxx~}%
    \fi}
\gdef\xxx{\@ifnextchar[\xxx@lab\xxx@nolab}
\long\gdef\xxx@lab[#1]#2{{\bf [\xxxmark \textcolor{red}{#2} ---{\sc #1}]}}
\long\gdef\xxx@nolab#1{{\bf [\xxxmark \textcolor{red}{#1}]}}
\ifnum\isReadyToSubmit=1
    \long\gdef\xxx@lab[#1]#2{}\long\gdef\xxx@nolab#1{}
\fi
}

{\makeatletter
\gdef\edit{\@ifnextchar[\edit@lab\edit@nolab}
\long\gdef\edit@lab[#1]#2{[\textcolor{red}{#2} ---{\sc #1}]}
\long\gdef\edit@nolab#1{[\textcolor{red}{#1}]}
\ifnum\isReadyToSubmit=1
    \long\gdef\edit@lab[#1]#2{[#2]}
\fi
}

\newcommand{\ignore}[1]{}

\definecolor{grey}{rgb}{0.5,0.5,0.5}

\algdef{SE}[VARIABLES]{Variables}{EndVariables}
{\algorithmicvariables}{}
\algnewcommand{\algorithmicvariables}{\textbf{global variables}}

\def\ie{{i.e.}\xspace}
\def\eg{{e.g.}\xspace}
\def\etal{{et al.}\xspace}

\def\overflow_relevance{overflow relevance\xspace}
\def\RK{privacy knapsack\xspace}
\def\RDP{RDP\xspace}
\def\TDP{traditional DP\xspace}

\def\OPT{Optimal\xspace}

\def\PrivacyLoss{\textrm{PrivacyLoss}}

\def\ourmethod{DPack\xspace}

\def\ourimprovementmicro{1.0--2.6$\times$\xspace} 
\def\ourimprovementalibaba{1.3--1.7$\times$\xspace}

\def\pmax{w^{\max}}
\def\pmaxhat{{\hat w}^{\max}}
\def\alphamaxhat{{\hat \alpha}^{\max}}

\newtheorem{theorem}{Theorem}

\newtheorem{property}[theorem]{Property}

\def\E{\mathbb{E}}
\def\R{\mathbb{R}}
\newcommand{\N}{{\mathbb N}}

\def\cA{\mathcal{A}}
\def\cD{\mathcal{D}}

\def\cS{\mathcal{S}}
\def\cY{\mathcal{Y}}

\begin{document}

\date{}  %

\title{\ourmethod: Efficiency-Oriented Privacy Budget Scheduling}

\author{Pierre Tholoniat*}
\affiliation{
    \institution{Columbia University}
    \city{New York City}
    \state{NY}
    \country{USA}
}

\author{Kelly Kostopoulou*}
\affiliation{
    \institution{Columbia University}
    \city{New York City}
    \state{NY}
    \country{USA}
}

\author{Mosharaf Chowdhury}
\affiliation{
    \institution{University of Michigan}
    \city{Ann Arbor}
    \state{MI}
    \country{USA}
}

\author{Asaf Cidon}
\affiliation{
    \institution{Columbia University}
    \city{New York City}
    \state{NY}
    \country{USA}
}
\author{Roxana Geambasu}
\affiliation{
    \institution{Columbia University}
    \city{New York City}
    \state{NY}
    \country{USA}
}
\author{Mathias L\'ecuyer}
\affiliation{
    \institution{University of British Columbia}
    \city{Vancouver}
    \state{BC}
    \country{Canada}
}

\author{Junfeng Yang}
\affiliation{
    \institution{Columbia University}
    \city{New York City}
    \state{NY}
    \country{USA}
}

\begin{abstract}
	Machine learning (ML) models can leak information about users, and differential privacy (DP) provides a rigorous way to bound that leakage under a given budget.  This DP budget can be regarded as a new type of computing resource in workloads of multiple ML models training on user data.  Once it is used, the DP budget is forever consumed. Therefore, it is crucial to allocate it most efficiently to train as many models as possible.  This paper presents a scheduler for the privacy resources that optimizes for efficiency.  We formulate privacy scheduling as a new type of multidimensional knapsack problem, called {\em \RK}, which maximizes DP budget efficiency. We show that \RK is NP-hard, hence practical algorithms are necessarily approximate.  We develop an approximation algorithm for privacy knapsack, {\em \ourmethod}, and evaluate it on microbenchmarks and on a new, synthetic private-ML workload we developed from the Alibaba ML cluster trace.  We show that \ourmethod: (1) often approaches the efficiency-optimal schedule, (2) consistently schedules more tasks compared to a state-of-the-art privacy scheduling algorithm that focused on fairness instead of efficiency (\ourimprovementalibaba in Alibaba, \ourimprovementmicro in microbenchmarks), but (3) sacrifices some level of fairness for efficiency. Using \ourmethod, DP ML operators should be able to train more models on the same amount of user data while offering the same privacy guarantee to their users.
\end{abstract}

    

\maketitle
\def\thefootnote{*}\footnotetext{These authors contributed equally to this work.
}\def\thefootnote{\arabic{footnote}}

\thispagestyle{plain}
\pagestyle{plain} %

\section{Introduction}
\label{sec:introduction}

Machine learning (ML) models are consuming an essential resource -- {\em user privacy} -- but they are typically not accounting for or bounding this consumption.
A large company may train thousands of models over user data per week, continuously updating its models as it collects new data.
Some of the models may be released to mobile devices or distributed globally to speed up inference.
Unfortunately, there is increasing evidence that ML models can reveal specific entries from their original training sets~\cite{dwork2017exposed,carlini2018theSecretSharer,shokri2017membership,carlini2020extracting, nasr2023scalableextractiontrainingdata}, both through parameters and predictions, thereby potentially leaking user data to adversaries.
Intuitively, the more one learns from aggregate user data, the more one should expect to also learn (and hence leak) about individual users whose data is used.
This intuition has been proven formally for simple statistics~\cite{dinurNissim2003revealing} and repeatedly demonstrated experimentally for ML models~\cite{carlini2018theSecretSharer,shokri2017membership, nasr2023scalableextractiontrainingdata}.
Therefore, user privacy can be viewed as a {\em resource} that is consumed by tasks in an ML workload, and whose consumption should be accounted for and bounded to limit data leakage risk.

Differential privacy (DP)~\cite{Dwork:2006:CNS:2180286.2180305} provides a rigorous way to define the privacy resource, and to account for it across multiple computations or tasks, be they ML model training tasks or statistic calculations.
DP randomizes a computation over a dataset (\eg training an ML model or computing a statistic) to bound the leakage of entries in the dataset through the output of the computation~\cite{privatekube}. Each DP computation increases this bound on data leakage, consuming some of the data's {\em privacy budget}, a pre-set quantity that should never be exceeded to maintain the privacy guarantee. In workloads with a large number of tasks that continuously train models on a private corpus or stream, the data's privacy budget is a very scarce resource that must be efficiently allocated to enable the execution of as many tasks as possible.

In our prior work~\cite{privatekube,sage}, we began exploring how to expose data privacy as a new {\em computing resource} that is inherently being consumed by the tasks in an ML cluster and which must therefore be allocated and managed by the cluster's resource manager similarly to how other, more traditional computing resources -- CPU, GPU, and RAM -- are managed.  Other researchers proposed Cohere~\cite{cohere} an alternative approach for treating privacy as a computing resource.  A common conclusion of these prior works is that because the privacy resource behaves differently from traditional computing resources (\eg it is finite), scheduling it requires new algorithms.
To this end, we proposed DPF~\cite{privatekube}, the first scheduling algorithm for the privacy resource, which adapted the well-known dominant resource fairness (DRF) algorithm to the privacy resource.  Our focus was on {\em fairness} as the key objective for our algorithm design. DPF guarantees a form of max-min fairness for the privacy budget when multiple tasks compete for it.

Unfortunately, as is often the case in scheduling~\cite{carbyne,hug,joe2013multiresource,fair-allocation,parkes2015beyond,tetris}, fairness can come at the cost of allocation {\em efficiency}, measured as the total number of tasks that are allocated over a unit of time.
For privacy, we find that this inefficiency is especially evident in workloads that exhibit a high degree of \emph{heterogeneity} either in the data segments they request (\eg, a workload containing tasks that run on data collected from different time ranges), or in the types of tasks they contain (\eg a workload mixing different types of statistics and ML algorithms). In such cases, we show that a scheduler that optimizes for efficiency rather than fairness can schedule up to 2.6$\times$ more tasks than DPF for the same privacy budget.

In this paper, we explore the first practical {\em efficiency-oriented privacy schedulers}, which aim to maximize the number of scheduled tasks, or the total utility of scheduled tasks if tasks are assigned utility weights (\S\ref{sec:efficiency-algorithms}).
We first introduce a new formulation of the DP scheduling problem, which optimizes for efficiency, and show that it maps to the NP-hard multidimensional knapsack problem, requiring practical approximations to solve in practice.
We demonstrate that (1) our prior DPF algorithm, which optimizes for fairness, can be seen as an inefficient heuristic to solve this problem, and that (2) a better heuristic for multidimensional knapsack yields more efficient DP scheduling.
We then show that instantiating the privacy scheduling problem to R\'enyi DP (RDP) accounting, a state-of-the-art, efficient DP accounting mechanism, introduces a new dimension with unusual semantics to the scheduling problem.
To support this new dimension, we define a new knapsack problem that we call the {\em \RK}, which we show is also NP-hard.
Finally, we propose a new \RDP-aware heuristic for the \RK, instantiate it into a new scheduling algorithm called {\em \ourmethod}, provide a formal analysis of its approximation properties, and discuss when one should expect to see significant efficiency gains from it (\S\ref{sec:applicability}).

We implement \ourmethod in a Kubernetes-based orchestrator for data privacy~\cite{privatekube} and an easily-configurable simulator~(\S\ref{sec:implementation}).
Using both microbenchmarks and a new, synthetic, DP-ML workload we developed from the Alibaba's ML cluster trace~\cite{alibaba-gpu}, we compare \ourmethod to DPF, the optimal \RK solver, and first-come-first-serve (FCFS) (\S\ref{sec:evaluation}).
\ourmethod schedules significantly more tasks than DPF (\ourimprovementalibaba in Alibaba and \ourimprovementmicro in microbenchmarks), and closely tracks the optimal solution, at least up to a small number of blocks and tasks where it is feasible for us to obtain the optimal solution. \ourmethod on Kubernetes can scale to thousands of tasks, and incurs a relatively modest scheduler runtime overhead.  Still, by focusing on efficiency, \ourmethod sacrifices some level of fairness compared to DPF: in the Alibaba workload, DPF is able to schedule 90\% of tasks that request less or equal than their privacy budget ``fair-share'', while \ourmethod schedules only 60\% of such tasks.
This is inevitable given the rather fundamental tradeoff between efficiency and fairness in scheduling.
Our work thus fills in an important gap on algorithms that prioritize efficiency over fairness, as we believe will be desirable given the scarcity of this essential new resource in ML systems, user privacy.

This paper is organized as follows.  \S\ref{sec:background} provides background on the threat model we are addressing, DP, and prior work on DP scheduling.  Much of this section builds upon our prior papers in this space~\cite{privatekube, sage}, so there is considerable redundancy in the statements with those papers', which we include for the purposes of making this paper self-contained.  \S\ref{sec:efficiency-algorithms} begins our main contributions in this work, consisting of the definitions and hardness properties of the efficiency-oriented DP scheduling problem, its adaptation for \RDP, and the \ourmethod algorithm we propose for both efficient and practical DP resource scheduling.  \S\ref{sec:applicability} describes the applicability of our approach, highlighting cases when \ourmethod is likely to give substantial efficiency benefit compared to DPF, as well as cases when it will not do so.  \S\ref{sec:implementation} presents our implementation of \ourmethod, while \S\ref{sec:evaluation} provides our evaluation. Finally, \S\ref{sec:related-work} reviews related works and \S\ref{sec:conclusion} concludes.  We make our prototype and experimental code available at \url{https://github.com/columbia/dpack}.

\section{Background}
\label{sec:background}

\subsection{Threat Model}
\label{sec:background:threat-model}

We adopt the same threat model as in our prior work~\cite{privatekube}.  We are concerned with the sensitive data exposure that may occur when pushing models trained over user data to untrusted locations, such as end-user devices or inference servers all around the world.  We operate under a centralized-DP model: a trusted curator collects and stores all user data and executes tasks, which consist of ML training procedures or pipelines that are explicitly programmed to satisfy a particular $(\epsilon, \delta)$-DP guarantee.  We trust that the curator and the programmers of the tasks are not malicious and will not want to inspect, steal, or sniff the data.
However, we do not trust the recipients of results released by the system, or the locations in which they are stored.  
Those results may be statistical aggregates, ML model predictions, or entire ML models. Accessing them may allow malicious activities that compromise sensitive personal information. We impose DP guarantees across all the processes that generate them.
{\em Membership inference} attacks \cite{backes2016membership, dwork2015robustTraceability, homer2008resolving, shokri2017membership} allow the adversary to infer whether an individual is in the data used to generate the output. {\em Data reconstruction attacks} \cite{carlini2018theSecretSharer, dinurNissim2003revealing, dwork2017exposed} allow the adversary to infer sensitive attributes about individuals that exist in this data.
We tackle both types of attack.

Our focus is not on single models or statistics, released once, but rather on {\em workloads of many models or statistics}, trained or updated periodically over windows of data from user streams.
For example, a company may train an auto-complete model daily or weekly to incorporate new data from an email stream, distributing the updated models to mobile devices for fast prediction.
Moreover, the company may use the same email stream to periodically train and disseminate multiple types of models, for example for recommendations, spam detection, and ad targeting. This creates ample opportunities for an adversary to collect models and perform privacy attacks to siphon personal data.
To prevent such attacks, our goal is to {\em maintain a global ($\epsilon^{G}$, $\delta^{G}$)-DP guarantee over the entire workload consisting of many tasks}.

\subsection{DP Background}
\label{sec:background:dp}

We present background on DP theory that is necessary to understand our scheduling algorithm. DP addresses both membership inference and data reconstruction attacks~\cite{shokri2017membership,dwork2017exposed,carlini2018theSecretSharer,236254}.
Intuitively, both attacks work by finding data points (which can range from individual events to entire users) that make the observed model more likely: if those points were in the training set, the likelihood of the observed model increases. DP prevents these attacks by ensuring that no specific data point can drastically increase the likelihood of the model produced by the training procedure.

DP randomizes a computation over a dataset (such as the training of  ML model) to bound a quantity called {\em privacy loss}, defined as some measure of the change in the distribution over the outputs of the randomized computation incurred when a single data point is added to or removed from the input dataset.
Privacy loss is a formalization of what one might colloquially call ``leakage'' through a model.
Satisfying DP means bounding privacy loss by some fixed, parameterized value, $\epsilon > 0$, which is called {\em privacy budget}. This bound is enforced through the virtue of the randomness (often called noise) added into the computation.  There are multiple ways to define privacy loss, corresponding to various ways to define the distance between two output distributions.  These different privacy loss definitions lead to different DP definitions, each with different interpretations, strengths and weaknesses.  We review two DP definitions here.

\heading{Traditional differential privacy ($\epsilon$-DP and $(\epsilon, \delta)$-DP).}
The original definition proposed by Dwork, et al.~\cite{Dwork:2006:CNS:2180286.2180305} defines privacy loss as follows.
Given a randomized algorithm, $\cA : \cD \rightarrow \cY$, for any datasets $\cD, \cD'$ that differ in one entry (called {\em neighboring datasets}) and for any output $y \in \cY$:
\begin{equation}\label{eq:dp_privacy_loss}
    \textrm{\PrivacyLoss}(y, \cD, \cD') = \log\Big(\frac{P(\cA(\cD) = y)}{P(\cA(\cD') = y)}\Big) .
    \vspace{-.2cm}
\end{equation}
The traditional, pure $\epsilon$-DP definition requires an algorithm to satisfy $|\PrivacyLoss(y, \cD, \cD')| \leq \epsilon$ for any $y$, $\cD$, $\cD'$ as above.
A variation of this definition, popularly used in ML, is $(\epsilon, \delta)$-DP: for $\delta \in [0,1)$, it requires an algorithm to satisfy 
$P(\cA(\cD) \in \cS) \leq \exp(\epsilon) P(\cA(\cD') \in \cS) +\delta$ for all $\cS \subseteq \text{Range}(\cA)$, for each neighboring $\cD, \cD'$.

These traditional DP definitions have the strength of being relatively interpretable: for a small value of $\epsilon$ (e.g., $\epsilon \le 1$), $\epsilon$-DP can be interpreted as a guarantee that an attacker who inspects the output of an $\epsilon$-DP computation will not learn anything new with confidence about any entry in the training set that they would not otherwise learn if the entry were not in the training set~\cite{vadhan2017complexity}.
Similarly, for small $\delta$ (e.g., $\delta<\frac{1}{n^2}$ for dataset size $n$), $(\epsilon, \delta)$-DP guarantee is roughly a high-probability $\epsilon$-DP guarantee.
The advantage of $(\epsilon, \delta)$-DP is support for a richer set of randomization mechanisms, such as adding noise from a Gaussian distribution, which pure DP cannot, and which often provide better privacy-utility tradeoffs.
That is why $(\epsilon, \delta)$-DP is the reference privacy definition for DP ML.

\heading{R\'enyi DP Accounting ($(\alpha, \epsilon)$-\RDP).}
More recent DP definitions define privacy loss differently, usually sacrificing interpretability for tighter analysis of randomization mechanisms and how they compose with each other, yielding even better privacy-utility tradeoffs, especially in DP ML.  A state-of-the-art definition is \RDP~\cite{8049725}, which has been adopted internally by most DP ML platforms~\cite{google-dp-rdp,tensorflow-privacy,opacus}.  
Instead of defining the privacy loss based on probability ratios as traditional DP does, \RDP defines it in terms of the R\'enyi divergence, a particular distance between the distributions over all possible outcomes for $\cA(\cD)$ and $\cA(\cD')$.
R\'enyi divergence has a parameter, $\alpha > 1$, called {\em order}:
\begin{equation*}\label{eq:rdp_privacy_loss}
    \textrm{\PrivacyLoss}_\alpha(\cD, \cD') = \frac{1}{\alpha -1} \log \underset{{\footnotesize y \sim \cA(\cD)}}{\E} \Big( \frac{P(\cA(\cD)=y)}{P(\cA(\cD')=y)} \Big)^\alpha.
\end{equation*}
As before, $(\alpha, \epsilon)$-\RDP requires that $|\PrivacyLoss_\alpha(\cD, \cD')| \leq \epsilon$ for any datasets $\cD, \cD'$ differing in one entry.

\RDP is less interpretable than \TDP due to the complexity of R\'enyi divergence.
However, one can always translate from $(\epsilon, \alpha)$-\RDP to $(\epsilon_{DP}, \delta)$-DP~\cite{8049725} for any appropriately ranged values of $\alpha$, $\epsilon$, and $\delta$:
\begin{equation}\label{eq:rdp_to_dp}
    \epsilon_{DP} = \epsilon + \frac{\log(1/\delta)}{\alpha-1} .
\end{equation}

\RDP's greatest advantage over \TDP\ -- and the reason for its recent adoption by most major DP ML platforms as well as for our special consideration of it in this paper -- is its support for {\em both efficient and convenient composition}.  
All successful DP definitions are closed under composition; i.e., running multiple DP computations satisfies the DP definition, albeit with a worse $\epsilon$ parameter.  
However, whereas with \TDP, composing $m$ mechanisms degrades the global guarantee linearly with $m$, with \RDP, the global guarantee degrades with $\sqrt{m}$ when applying composition followed by conversion to \TDP through Eq.~\ref{eq:rdp_to_dp}.
\RDP's tighter analysis can allow composition of more DP computations with the same $\epsilon$ guarantees; the advantage is particularly significant with a large $m$.

Since popular DP ML algorithms, such as DP SGD, consist of tens of thousands iterations of the same rudimentary DP computation (computing one gradient step over a sample batch), they require the most efficient composition accounting method.
This is why most DP ML platforms internally operate on RDP to compose across training steps and then translate the cumulative \RDP guarantee into \TDP (with Eq.~\ref{eq:rdp_to_dp}) to provide an interpretable privacy semantic externally.
Similarly, since our goal is to develop {\em efficient scheduling algorithms} -- that pack as many DP ML tasks as possible onto a fixed privacy budget -- it is incumbent on us to consider \RDP accounting in our scheduling formulations.\footnote{We considered, and discarded, advanced composition for \TDP, which is also efficient but involves complex arithmetic that is untenable to incorporate in a scheduler~\cite{complexity_advanced_composition}.}
We do so in a similar way: internally, some of the algorithms we propose use \RDP accounting (albeit to compose across ML training tasks, {\em not} across gradient steps within a task) but externally we will always expose a \TDP guarantee.
As it turns out, operating on \RDP internally creates interesting challenges for scheduling, about which we discuss in \S\ref{sec:offline-rdp-composition}.

\subsection{Privacy Scheduling Background}
\label{sec:background:dpf}

In a recent line of work~\cite{sage, privatekube}, we have argued for the global privacy budget to be managed as a new type of computing resource in workloads operating on user data: its use should be tracked and carefully allocated to competing tasks.
We adopt the same focus on ML platforms for continuous training on user data streams, such as Tensorflow-Extended (TFX), and build on the same basic operational model~\cite{sage} and key abstractions and algorithms~\cite{privatekube} for monitoring and allocating privacy in DP versions of these platforms.
The operational model is as follows.
Similar to TFX, the user data stream is split into multiple non-overlapping {\em blocks} (called {\em spans} in TFX~\cite{spans}), for example by time, with new blocks being added over time.
Blocks can also correspond to partitions given by SQL \texttt{GROUP BY} statements over public keys, such as in Google's DP SQL system \cite{zetasql} or in the DP library used for the U.S. Census \cite{tumult}.
There are multiple tasks, dynamically arriving over time, that request to compute (e.g., train ML models) on subsets of the blocks, such as the most recent $N$ blocks.
The company owning the data wants to enforce a global \TDP guarantee, $(\epsilon^G,\delta^G)$-DP, that cannot be exceeded across these tasks.
Each data block is associated with a global privacy budget (fixed a priori), which is consumed as DP tasks compute on that block until it is depleted.

In Luo \etal~\cite{privatekube}, we incorporated {\em privacy blocks}, i.e., data blocks with privacy budget, as a new compute resource into Kubernetes, to allocate privacy budget from these blocks to tasks that request them.
The resulting system, which is a drop-in extension of Kubernetes, is called {\em PrivateKube}.
To request privacy budget from a privacy block, a task $i$ sets a demand vector ($d_{i}$) of length $m$, equal to the number of blocks in the system. The demand vector specifies the privacy budget that task $i$ requests for each individual block in the system (with a zero demand for blocks that it is not requesting).  If task $i$ is allocated, then its demand vector is consumed from the blocks' privacy budgets.  When a block's privacy budget reaches zero, no more tasks can be allocated for that block and the block is removed.  This ensures that a block of user data will not be used to extract so much information that it risks leaking information about the users.
In this sense, each privacy block is a \emph{non-replenishable} or finite resource.
It is therefore important to carefully allocate budget from privacy blocks across tasks, so as to pack as many tasks as possible onto the blocks available at any time.
That's the goal of {\em efficiency-oriented privacy scheduling} and it is in contrast (and as we shall see, at odds) with fairness-oriented scheduling, which we previously explored in PrivateKube with an algorithm called {\em DPF} (Dominating Privacy-block Fairness).  We defer a description of DPF and the tradeoffs between fairness and efficiency in privacy scheduling until after we have formulated the efficiency-oriented privacy scheduling problem in what follows.

\section{Efficiency-Oriented Privacy Scheduling}
\label{sec:efficiency-algorithms}

A key contribution of this work is the formalization of the efficiency-oriented DP scheduling problem.
We first develop an {\em offline} version of this problem, in which the entire workload is assumed to be fixed and known a priori, and study efficient DP scheduling under traditional DP and basic composition (\S\ref{sec:offline-basic-composition}).
We show that offline DP scheduling maps to the NP-hard multidimensional knapsack problem, requiring practical approximations to solve in practice.
Describing how our previous DPF algorithm works, we show that it can be seen as an inefficient heuristic for the efficiency-oriented scheduling problem, albeit one that has fairness guarantees.
We then show that a better heuristic yields more efficient DP scheduling with multiple data blocks.
In \S\ref{sec:offline-rdp-composition} we move onto a more complex \RDP formulation of the efficiency-oriented allocation problem, but one that has the potential to boost efficiency significantly compared to \TDP thanks to \RDP's composition benefits.  We prove the new RDP formulation as also NP-hard and develop a second, \RDP-aware heuristic that leverages some unusual characteristics of this problem.
In \S\ref{sec:greedy-algo}, we describe {\em \ourmethod}, our proposed efficiency-oriented scheduling algorithm that incorporates both of our heuristics and in special settings can be shown to be a proper approximation of the efficiency-optimal solution to the RDP privacy knapsack problem.
Finally, in \S\ref{sec:online} we adapt \ourmethod to the online case.

\subsection{Efficient Scheduling with Traditional DP}
\label{sec:offline-basic-composition}

We define the {\em global efficiency} of a scheduling algorithm as either the number of scheduled tasks or, more generally, the sum of {\em weights} $w_i$ of scheduled tasks, for cases when different tasks have different utilities (a.k.a. profits or weights) to the organization.
When the goal is to optimize global efficiency, we can model privacy budget scheduling in a multi-block system such as TFX as a multidimensional knapsack problem.  First, recall that traditional DP composes, in its simplest form, using an additive arithmetic: the composition of two $(\epsilon_1, \delta_1)$-DP and $(\epsilon_2, \delta_2)$-DP tasks is $(\epsilon_1+\epsilon_2, \delta_1+\delta_2)$-DP.  In this paper we assume $\delta$ is extremely small (as it should always be, since it is a failure probability of the pure DP guarantee), hence we ignore the additive effects on the $\delta$ parameters and instead focus on the additive effects of the $\epsilon$ parameters, which are typically many orders of magnitude larger than the $\delta$ parameters.

\heading{Knapsack problem formulation.}
Consider a fixed number of $n$ tasks ($t_1, \dots, t_n$) that need to be scheduled over $m$ blocks, each with $c_j$ remaining capacity. Each task has a demand vector $d_{ij}$, which represents the $\epsilon$ demand by task $i$ for block $j$, and a weight $w_i$ if it is successfully scheduled (when $w_i$ is equal across all tasks, the problem is to maximize the number of scheduled tasks).
We can formulate this problem as the standard multidimensional knapsack problem \cite{knapsack_problems_textbook}, where $x_i$ are binary variables:
\begin{equation}\label{eq:knapsack-problem}
    \underset{x_i \in \{0,1\}}{\max} \sum_{i=1}^n w_ix_i \text{ subject to } \forall j \in [m]: \sum_{i=1}^n d_{ij} x_i \le c_{j} .
\end{equation}

W.l.o.g., we assume there is not enough budget to schedule all tasks:
$\forall j \in [m]: \sum_{i=1}^n d_{ij} > c_{j} .$
Otherwise, the knapsack problem is trivial to solve. If some blocks have enough budget but not others, we can set the blocks with enough budget aside, solve the problem only on the blocks with contention, and incorporate the remaining blocks at the end.

\begin{figure}[t]
    \centering
    \subfigure[Inefficient allocation with DPF]{
        \includegraphics[width=0.43\linewidth]{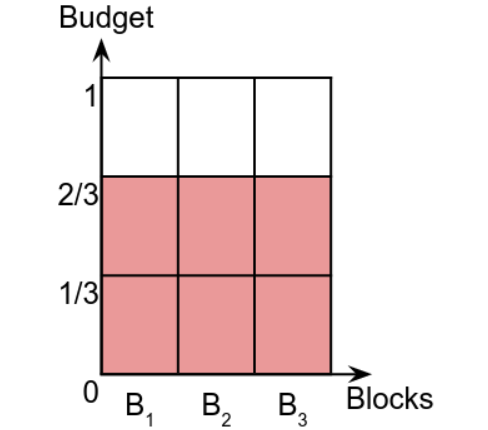}
        \label{fig:multiblock_fairness}
    }
    \subfigure[Efficient allocation]{
        \includegraphics[width=0.46\linewidth]{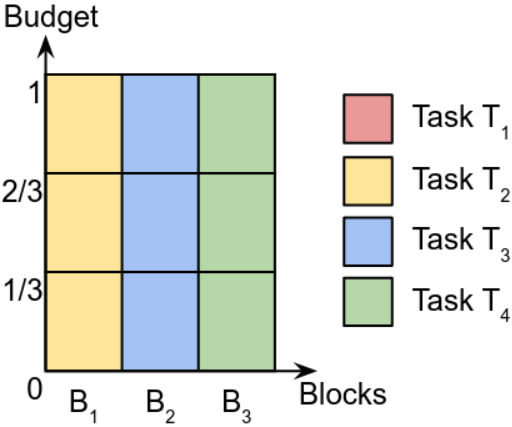}
        \label{fig:multiblock_packing}
    }
    \caption{\small {\bf Example of allocations with basic DP accounting.}
        Task $T_1$ requests privacy budget from 3 blocks, $B_1, B_2, B_3$.
        Tasks $T_2, T_3, T_4$ request slightly more privacy budget, but each one from one distinct block: $B_1, B_2, B_3$, respectively.
        In \ref{fig:multiblock_fairness}, DPF sorts these tasks based on their dominant shares: $T_1$ first (because its dominant share is lower, even though it demands budget from all the blocks), then $T_2, T_3, T_4$ in arbitrary order. After $T_1$ is scheduled, there is no more budget for other tasks.
        Meanwhile, in \ref{fig:multiblock_packing} an efficient scheduler can allocate 3 tasks.
    }
    \label{fig:dpf_multi_block_killer}
\end{figure}

\heading{The need for heuristics.}
The multidimensional knapsack problem is known to be NP-hard \cite{knapsack_problems_textbook},
so DP scheduling cannot be solved exactly, even in the offline case.
There exist some general-purpose polynomial approximations for this problem, but they are exponential in the approximation parameter and become prohibitive for large numbers of dimensions (for us, many blocks).
In \S\ref{sec:eval-offline-microbenchmark}, we show that the Gurobi~\cite{gurobi} solver quickly becomes intractable with just $7$ blocks!

A standard approach to practically solve knapsack problems is to develop specialized approximations for a specific domain of the problem, typically using a {\em greedy algorithm}
that sorts tasks according to a {\em task efficiency metric} (denoted $e_i$), and then allocates tasks in order, starting from the highest-efficiency tasks, until the algorithm cannot pack any new tasks~\cite{knapsack_problems_textbook}.
In such algorithms, the main challenge is coming up with good task efficiency metrics that leverage domain characteristics to meaningfully approximate the optimal solution while remaining practical in terms of runtime.

\heading{Inefficiencies under DPF, seen as a scheduling heuristic.}
Turns out we can model DPF -- our previous, fairness-oriented algorithm and still the state-of-the-art privacy scheduling algorithm -- as a {\em greedy heuristic for \RK}.
DPF schedules tasks with the smallest dominant share ($\max_{j} \frac{d_{ij}}{c_{j}}$) first.
Folding in task weights, this becomes equivalent to a greedy algorithm with an efficiency metric defined as:
$e_i := \frac{w_i}{\max_{j} \frac{d_{ij}}{c_{j}}}$.
Unfortunately, given this efficiency metric, DPF can stray arbitrarily far from the optimal even in simple cases.
The reason lies in the maxima over $j$, which is crucial to ensure the fair distribution of DP budget, but causes DPF to ignore multidimensionality in data blocks.
\F\ref{fig:dpf_multi_block_killer} gives an example using \TDP and a workload of 4 tasks.
DPF sorts tasks by dominant share and schedules only one task.
Meanwhile, a better efficiency metric would consider the ``area'' of a task's demand, thereby sorting tasks $T_2, T_3$ and $T_4$ before $T_1$, resulting in 3 tasks getting scheduled.
Thus, DPF, despite its compelling weighted fairness guarantees, is merely a greedy heuristic when it comes to optimizing for efficiency; it is not even a proper approximation of the efficiency-optimal allocation, as it can stray arbitrarily far from it.

\heading{Area-based metric for efficient scheduling over blocks.}
We take inspiration from single-dimensional knapsacks, in which the efficiency $e_i$ of task $i$ is usually defined as the task's weight-to-demand ratio: $e_i := w_i/d_i$.
A natural extension to multiple blocks uses a known heuristic for multidimensional knapsacks~\cite{vector_bin_packing} to capture the entire demand of a task:
\begin{equation}\label{eq:simple_task_efficiency}
    e_i := \frac{w_i}{\sum_{j} \frac{d_{ij}}{c_{j}}} ,
\end{equation}
\noindent where $\frac{d_{ij}}{c_{j}}$ is task $i$'s DP budget demand for block $j$, normalized by the remaining capacity of block $j$.
This normalization is important to express the scarcity of a demanded resource.
Unlike the DPF fair scheduling metric, Eq.~\ref{eq:simple_task_efficiency} considers the entire ``area'' of a task's demand to compute its efficiency, addressing the inefficiency from \F\ref{fig:dpf_multi_block_killer}.
A task requesting a large budget across blocks is not scheduled even if its demand on any block (dominant share) is small.  As we shall see in experimental evaluation, this heuristic leads to more efficient scheduling than under DPF under \TDP.

\subsection{Efficient Scheduling Under \RDP Accounting}
\label{sec:offline-rdp-composition}

The above heuristic is satisfactory for \TDP accounting, but practitioners and state-of-the-art ML algorithms use the much more efficient \RDP accounting.
With \RDP, multiple bounds on the privacy loss can be computed, for various \RDP orders $\alpha$ (Eq. \ref{eq:rdp_privacy_loss}).
This yields an {\em \RDP order curve $\epsilon(\alpha)$} for that computation.
For instance, adding noise from a Gaussian with standard deviation $\sigma$ into a computation results in $\epsilon(\alpha) = \frac{\alpha}{2\sigma^2}$.
Other mechanisms, such as subsampled Gaussian (used in DP-SGD) or Laplace (used in simple statistics), induce other RDP curves.
These curves are highly non-linear and their shapes differ among each other. This makes it difficult to know analytically what the privacy loss function will look like when composing multiple computations with heterogeneous RDP curves.
For this reason, typically the RDP $\epsilon$ bound is computed on a few discrete $\alpha$ values ($\{1.5,1.75,2,2.5,3,4,5,6,8,16,32,64\}$~\cite{8049725}), on which the composition is performed.
Importantly, composition of $\epsilon$ parameters at each $\alpha$ value is still additive, a key element of \RDP's practicality.

\begin{figure}[t]
    \centering
    \subfigure[\RDP curves]{
        \includegraphics[width=.45\linewidth]{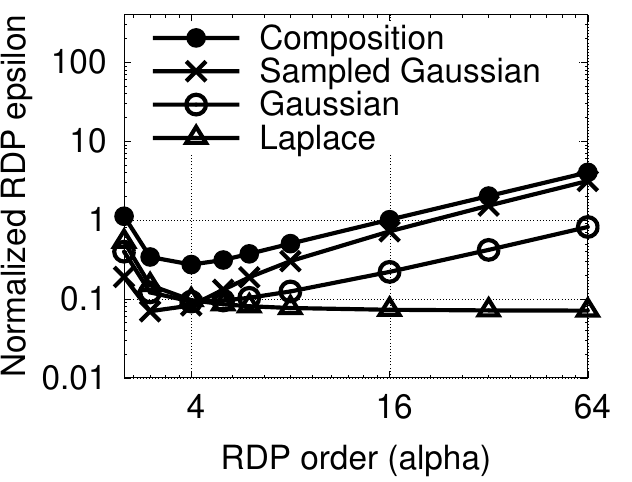}

        \label{fig:rdp_curve_examples}
    }
    \subfigure[DP translation]{
        \includegraphics[width=.41\linewidth]{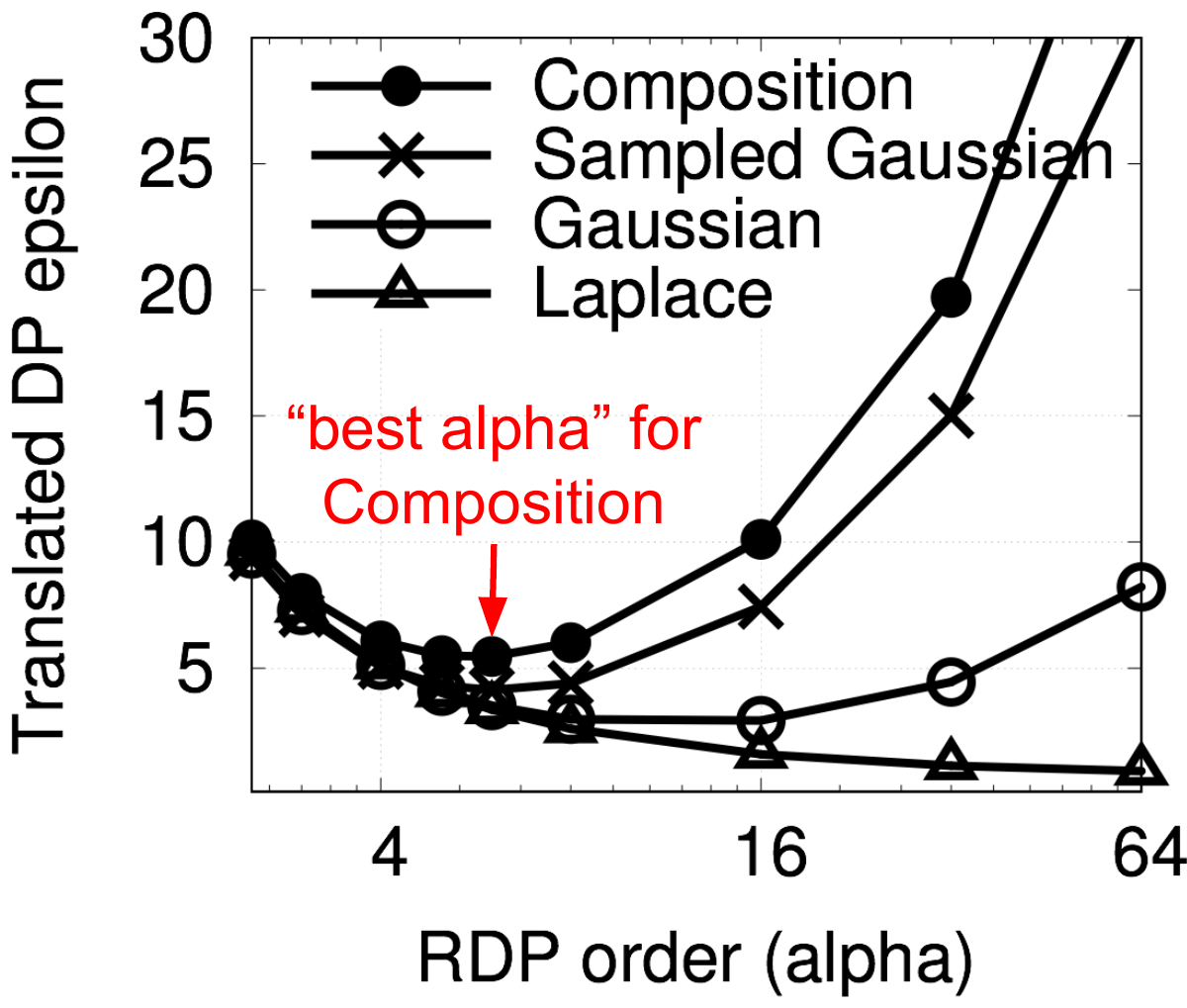}
        \label{fig:rdp_to_dp_curve_examples}
    }
    \caption{\small {\bf Example \RDP curves and DP translation.}
        (a) \RDP curves for Gaussian, subsampled Gaussian, and Laplace mechanisms, each with std-dev $\sigma=2$, plus their composition.
        (b) Translation to $(\epsilon_{DP}, 10^{-6})$-DP. The ``best'' (i.e., tightest) alpha differs among mechanisms.
        For composition, best is $\alpha=6$, giving $\epsilon_{DP}=5.5$.
    }
    \label{fig:rdp-curves}
\end{figure}

\F\ref{fig:rdp_curve_examples} shows \RDP curves for three example computations, each using a popular DP mechanism: the Gaussian would be used for a multidimensional statistic (a histogram); the subsampled Gaussian would be used in DP-SGD training; and Laplace would be used for a simple statistic (an average).  All these are plausible to co-exist as tasks in an ML/data analytics cluster.
These different computations exhibit different \RDP curves, with different orderings of the R\'enyi divergence bound at different $\alpha$'s. The subsampled Gaussian is tighter at lower $\alpha$ values; the Laplace is tighter for large $\alpha$'s.
The figure also shows the RDP curve for the composition of the three computations.

\F\ref{fig:rdp_to_dp_curve_examples} shows the translation of these four curves into \TDP (using Eq. \ref{eq:rdp_to_dp}).
For each computation, any value of $\alpha>1$ will translate into a different traditional $\epsilon$.
Some traditional $\epsilon$ translations are very loose, others are tighter, but they are all valid simultaneously.
Because of this, we can pick the $\alpha$ that gives us the best traditional $\epsilon$ guarantee and disregard the rest as loose bounds.
This {\em best alpha} differs from computation to computation: in our example, for the Gaussian it is $\alpha\approx16$; for the subsampled Gaussian $\alpha\approx6$; and for the Laplace $\alpha\ge 64$.
The best alpha for the composition of all three computations is $\alpha\approx6$, yielding $(\epsilon=5.5, \delta=10^{-6})$-DP.
If we were to analyze and compose the three computations directly in \TDP, we would obtain a looser global guarantee of $(\epsilon=7.8, \delta=10^{-6})$-DP. This gap grows fast with the number of computations.
Herein lies RDP's power,
but also a significant challenge when trying to allocate its privacy budget across competing computations.

Notice that when translating from \RDP to \TDP with Eq. \ref{eq:rdp_to_dp}, one chooses the most advantageous $\alpha$ for the final \TDP guarantee, ignoring all other \RDP orders.
This new $\alpha$ dimension therefore has a different semantic than the traditional multidimensional knapsack one.
Indeed, the traditional knapsack dimension semantic is that an allocation has to be within budget {\em along all dimensions}. This is a good fit for our block dimension, as we saw in \S\ref{sec:offline-basic-composition}.
Instead, an allocation is valid along the $\alpha$ dimension as long as the allocation is within budget for {\em at least one dimension}.
This creates opportunities for efficient scheduling, as the allocator can go over-budget for all but one $\alpha$ order. It also creates a new challenge, as the $\alpha$ order that will yield the most efficient allocation is unknown a priori and depends on the chosen combination of tasks.
Since the traditional multidimensional knapsack does not encode this new semantic, we define a new multidimensional knapsack problem for efficient \RDP scheduling.

\heading{The RDP privacy knapsack problem.} To accommodate \RDP, we need to modify the standard multidimensional knapsack problem to support alpha orders for each block and task demand. We express the capacity as $c_{j\alpha}$ (the available capacity of block $j$ on order $\alpha$), each demand vector as $d_{ij\alpha}$ (the demand of task $i$ on block $j$'s order $\alpha$), and require that the sum of the demands will be smaller or equal to the capacity for \emph{at least one of the alpha orders}.
We thus formulate the \RK as follows:

\begin{equation}\label{eq:privacy_knapsack}
\underset{x_i \in \{0,1\}}{\max} \sum_{i=1}^n w_ix_i \text{ subject to }\forall j \in [m], \exists \alpha \in A: \sum_{i=1}^n d_{ij\alpha} x_i \le c_{j\alpha} .
\end{equation}

We prove three properties of \RK (proofs in Appendix~\S\ref{sec:complexity-proofs}):
\begin{property}
    \label{prop:np-hard}
    The decision problem for the \RK problem is NP-hard.
\end{property}
\begin{property}
    \label{prop:single-block-fpta}
    In the single-block case, there is a fully polynomial time approximation scheme (FPTAS) for \RK, i.e., with $\pmax$ the highest possible global efficiency, for any $\eta>0$ we can find an allocation with global efficiency $\hat w$ such that $\pmax \leq (1+\eta)\hat w$, with a running time polynomial in $n$ and $1/\eta$.
\end{property}
\begin{property}
    \label{prop:no-fptas}
    For $m \ge 2$ blocks, there is no FPTAS for the \RK problem unless P=NP.
\end{property}
While Prop.~1 and~3 are disheartening (though perhaps unsurprising), Prop.~2 gives a glimmer of hope that at least for single-block instances, we can solve the problem tractably.
Indeed, as we shall see, this property is crucial for our solution.

\heading{DPF with multiple \RDP alpha orders.}
Fair scheduling with DPF for \RDP can once again be expressed as an ordering heuristic for the \RK, in which efficiency is defined as $e_i := \frac{w_i}{\max_{j\alpha} \frac{d_{ij\alpha}}{c_{j\alpha}}}$.
However, this approach is even more inefficient than under \TDP.

\begin{figure}[t]
    \centering
    \subfigure[Inefficient allocation with DPF]{
        \includegraphics[height=0.40\linewidth]{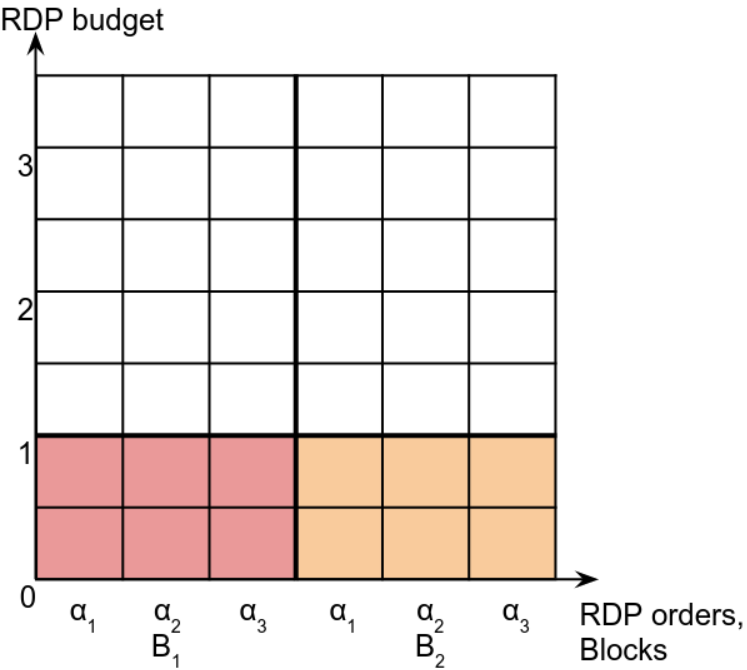}
        \label{fig:example_fairness_rdp}
    }
    \subfigure[Efficient allocation]{
        \includegraphics[height=0.40\linewidth]{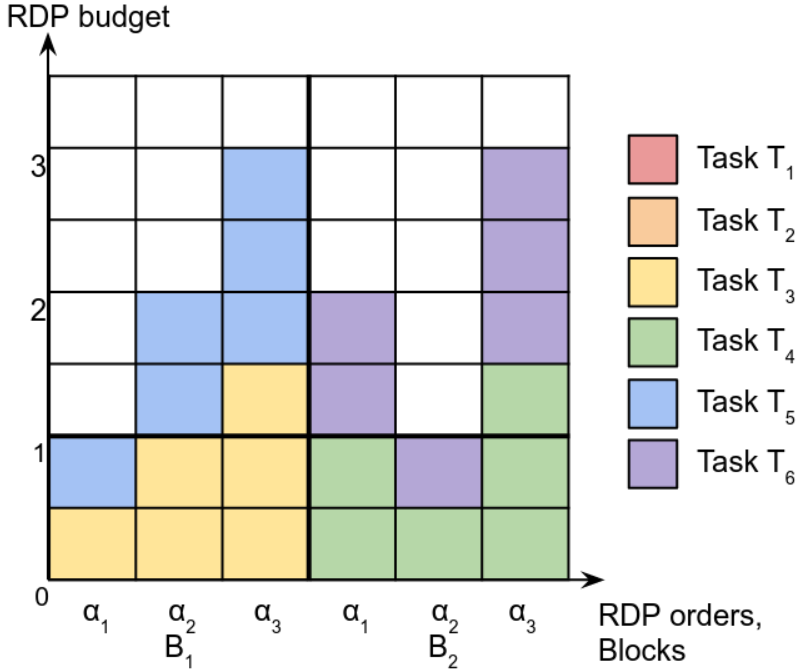}
        \label{fig:example_packing_rdp}
    }
    \caption{\small {\bf Example of allocations with \RDP accounting.}
        In \F\ref{fig:example_fairness_rdp}, DPF treats RDP orders like a regular resource and orders tasks by dominant share, allocating only 2 tasks in this example.
        Meanwhile, \F\ref{fig:example_packing_rdp} leverages the fact that only one order per block has to be below the capacity (here, $\alpha_1$ for block $B_1$ and $\alpha_2$ for block $B_2$). Tasks $T_3$ and $T_5$ have a large dominant share of $1.5$ but are efficient because they request only $0.5$ for $B_1$'s best alpha, $\alpha_1$.
    }
    \label{fig:example-rdp}
\end{figure}

In addition to the previous multi-block inefficiency (\S\ref{sec:offline-basic-composition}), this fair scheduling approach exhibits a new inefficiency under \RDP, regardless of the number of blocks it is invoked on (e.g., even if applied to non-block-based DP systems, such as DP SQL databases).
\F\ref{fig:example-rdp} gives an example using two blocks and a workload of 6 tasks, each requesting only 1 block. In \F\ref{fig:example_fairness_rdp}, DPF sorts tasks by the highest demands across all $\alpha$'s and allocates only 2 tasks.
A better efficiency metric would sort tasks by demands at the $\alpha$ value that can pack the most tasks (a.k.a., best alpha for composition), ultimately scheduling 4 tasks in \F\ref{fig:example_packing_rdp}.
Note that the best alpha is not necessarily the same for each block.

We conclude that an efficiency metric that simply takes the maximum of the dominant shares is neither efficient for scheduling multiple privacy blocks, nor for scheduling privacy budget in systems that use \RDP accounting.
However, a direct extension of our ``area based'' efficiency metric in Eq.~\ref{eq:simple_task_efficiency} does not appropriately handle \RDP alpha orders either, as it does not account for the specific semantic of the $\alpha$ order.
We next describe our new efficiency metric, that is optimized for efficiently scheduling tasks across multiple blocks and supports \RDP.

\subsection{\ourmethod Algorithm}
\label{sec:greedy-algo}

Intuitively, to support the ``at least one'' semantic of the $\alpha$ order from \RDP,
we need an efficiency metric that makes it less attractive to pack a task that consumes a lot of budget at what will ultimately be the {\em best alpha}, defined as the RDP order that packs the most tasks (or the most weight) while remaining under budget.
That best alpha is ultimately the only one for which the demands of tasks matter and hence should be the one used for computing an efficiency metric.
The challenge is that for workloads consisting of tasks with heterogeneous \RDP curves, the best alpha is not known a priori.
Our idea is to approximate it on a smaller set of \RDP curves, and to focus a task's efficiency metric on that best alpha as the only relevant dimension.
Recall from Prop.~\ref{prop:single-block-fpta} that in the single-block case, we can solve \RK with polynomial-time $\eta$-approximation for arbitrarily small $\eta>0$.
This means we can solve a single-block knapsack problem {\em separately for each block $j$} that determines the best alpha that will pack the most tasks (or maximal weight) among tasks requesting block $j$, taking only their request for that block into account.
We define the maximum utility for block $j$ and order $\alpha$ as $\pmax_{j\alpha} := \max_{x_i}{\sum_{i:d_{ij\alpha}>0}^n x_iw_i}$ subject to $\sum_i x_i d_{ij\alpha}  \le c_{j\alpha}$. We take $\pmaxhat_{j\alpha}$ a $\frac{2}{3}\eta$-approximation of $\pmax_{j\alpha}$ ($\frac{2}{3}$ is justified by proof below).
Based on this, we define the efficiency of task $i$ as:
\begin{equation}\label{eq:our_task_efficiency}
    e_i := \frac{w_i}{\sum_{j\alpha} (\frac{d_{ij\alpha}}{c_{j\alpha}} \text{ if } (\alpha == \arg\max_{\alpha'} \pmaxhat_{j\alpha'}) \text{ else 0}) }
\end{equation}
\begin{algorithm}[!t]
    \caption{{\bf \ourmethod Offline Algorithm}}
    \begin{algorithmic}
        \Variables
        \State tasks $i$, blocks $j$, RDP orders $\alpha$ capacities $c_{j\alpha}$
        \State approximation factor $\eta$, demands $d_{ij\alpha}$, weights $w_{i}$
        \EndVariables
        \Function{computeBestAlpha}{block $j$}
        \For{$\forall \alpha$}
        \State $\hat\pmax_{j\alpha} \gets$ \textproc{singleBlockKnapsack}($c_{\alpha}, d_{ij\alpha}, w_{i}, \frac{2}{3}\eta$)
        \EndFor
        \State return $\arg\max_{\alpha} \hat\pmax_{j\alpha}$
        \EndFunction

        \Function{computeEfficiency}{task $i$, best alphas $\alphamaxhat_{j}$}
        \State return $w_i / \sum_{j}( d_{ij\alphamaxhat_{j}} / c_{j\alphamaxhat_{j}})$

        \EndFunction

        \Function{CanRun}{task $i$}
        \State return $\forall j, \exists \alpha: \sum_{i'=1}^i d_{i'j\alpha} \le c_{j\alpha}$
        \EndFunction

        \Function{schedule}{tasks $i$}
        \For{$\forall j$}
        \State $\alphamaxhat_j \gets \Call{computeBestAlpha}{c_{j\alpha}, d_{ij\alpha}, w_{i}}$
        \EndFor
        \State $\textrm{sorted\_tasks} \gets$ $\textrm{tasks}$.sortBy(\textproc{computeEfficiency($\alphamaxhat_j$)})
        \For{$i$ in  $\textrm{sorted\_tasks}$}
        \If{\Call{CanRun}{$d_{ij\alpha}$}}
        \State Run task $i$, consuming the demanded budget
        \EndIf
        \EndFor
        \EndFunction
    \end{algorithmic}
    \label{alg:dpack}
\end{algorithm}

Alg.~\ref{alg:dpack} shows {\em \ourmethod}, our greedy approximation with the efficiency metric in Eq.~\ref{eq:our_task_efficiency}.
This algorithm addresses both of the problems we identified with DPF.
Moreover, we show that the manner in which \ourmethod handles \RDP is not just better than DPF in particular, but rather has two important generally desirable properties. First, \ourmethod reduces to the traditional multidimensional knapsack efficiency metric of Eq.~\ref{eq:simple_task_efficiency} when only one $\alpha$ exists, e.g. for \TDP:
\begin{property}
    \label{prop:tdp-support}
    If the dimension of $\alpha$ values is one (e.g., with \TDP), \ourmethod reduces to the traditional multidimensional knapsack heuristic from Eq. \ref{eq:simple_task_efficiency}.
\end{property}
\begin{proof}
    With one dimension, $\alpha = \arg\max_{\alpha'} \pmaxhat_{j\alpha'}$.
\end{proof}
Second, \ourmethod is a {\em guaranteed approximation} of the optimal in the specific cases when such an approximation is possible, the single-block case:
\begin{property}
    \label{prop:argmax_approximation}
    In the single-block case, \ourmethod is a ($\frac{1}{2} + \eta$)-approximation algorithm for \RK.
\end{property}
\begin{proof}
    Call $\hat \alpha := \arg\max_{\alpha'} \pmaxhat_{j\alpha'}$.
    By construction we have $\pmax_{j\hat\alpha} \leq (1+\frac{2}{3}\eta)\pmaxhat_{j\hat\alpha}$.
    In the single-block (index $j$) case, Eq.~\ref{eq:our_task_efficiency} means that tasks are greedily allocated by decreasing $\frac{w_i}{d_{ij\hat\alpha}}$, a well known $1/2$-approximation to the one dimensional knapsack problem~\cite{knapsack_problems_textbook}. Hence, $\pmax_{j\hat\alpha} \leq (1+\frac{2}{3}\eta) \pmaxhat_{j\hat\alpha} \leq (1+\frac{2}{3}\eta)(1+\frac{1}{2})\sum_{i=1}^n x_iw_i = (1+\frac{1}{2}+\eta)\sum_{i=1}^n x_iw_i$.
\end{proof}
Because of Prop.~\ref{prop:no-fptas}, a similar multi-block efficiency guarantee cannot be formulated (for \ourmethod as well as any other poly-time algorithm).
However, \S\ref{sec:evaluation} shows that in practice, \ourmethod performs close to the optimal solution of \RK in terms of global efficiency, yet it is a computationally cheap alternative to that intractable optimal solution.

\subsection{Adapting to the Online Case}
\label{sec:online}

In practice, new tasks and blocks arrive dynamically in a system such as TFX, motivating the need for an online scheduling algorithm.
We adapt our offline algorithm to the online case by scheduling a batch of tasks on the set of available blocks every $T$ units of time. To prevent expensive tasks from consuming all the budget prematurely, similar to DPF, we schedule each batch on a fraction of the total budget capacity: at each scheduling step we unlock an additional $1/N$ fraction of the block capacity.
More precisely, at each scheduling time $t = kT$, we execute Alg. \ref{alg:dpack} on the tasks and blocks currently in the system, but we replace block $j$'s capacity by:
\[ c_{j\alpha}^t = \frac{ \min(\lceil(t - t_j)/ T\rceil, N)}{N} \epsilon_{j\alpha} - \sum_{i' \in A_{t}} d_{i'j\alpha} ,\]

where $ \epsilon_{j\alpha}$ is the total capacity of block $j$ (computed from Prop. \ref{eq:rdp_to_dp}), $t_j$ is the arrival time of block $j$, $\lceil(t - t_j)/ T\rceil$ is the number of scheduling steps the block has witnessed so far (including the current step), and $A_{t}$ is the set of tasks previously allocated.

As with the offline algorithm, at the time of scheduling all the tasks are sorted by the scheduling algorithm. The scheduler tries to schedule tasks one-by-one in order.
Any tasks that did not get scheduled remain in the system until the next scheduling time, and any unused unlocked budget remains available for future tasks.
Users also specify a per-task timeout after which the task is evicted.
$T$ is a parameter of the system that controls how many tasks get batched (and delayed) before getting scheduled.
We evaluate its effect empirically in \F\ref{fig:batch-period-impact}, and show that beyond a reasonable batch size all algorithms we study are relatively insensitive to $T$.

Finally, to support a global $(\epsilon, \delta)$-DP guarantee for online tasks over continuous data streams, we use the data block composition introduced by Sage~\cite{sage, privatekube_arxiv}: each data block is associated with a privacy filter, a DP accounting mechanism enabling adaptive composition under a preset upper-bound on the privacy loss \cite{rogers2016privacy, feldman2021individual, adaptive_rdp}.
Each filter is initiated with $\epsilon, \delta$ for \TDP, or $\epsilon(\alpha) = \epsilon - \frac{\log(1/\delta)}{\alpha-1}$ for \RDP.
The \RDP initial value ensures that translating back to \TDP with Eq.~\ref{eq:rdp_to_dp} guarantees $(\epsilon, \delta)$-DP.
A task is granted if, and only if, all filters grant the request (all blocks have enough budget left).
This ensures the following property:
\begin{property}
  \vspace{-0.2cm}
  \label{prop:adaptive-dp}
  \ourmethod enforces $(\epsilon, \delta)$-DP over adaptively chosen computations and privacy demands $\epsilon_i(\alpha)$.
\end{property}
\begin{proof}
  We provide a proof sketch following the structure used in \cite[Theorem 4.2]{sage} for basic composition. Each task has an (adaptive) \RDP requirement for all blocks, with $\epsilon(\alpha) = 0$ for non-requested blocks.
  Each data block is associated with a privacy filter \cite[Algorithm 1]{adaptive_rdp}.
  A task runs if and only if all filters accept the task: applying \cite[Theorem 1]{adaptive_rdp} ensures $\epsilon(\alpha) = \epsilon - \frac{\log(1/\delta)}{\alpha-1}$-\RDP holds for each block. Applying Eq.~\ref{eq:rdp_to_dp} concludes the proof.
\end{proof}

\section{Applicability}
\label{sec:applicability}

It is worth reflecting on the characteristics of workloads under which \ourmethod provides the most benefit compared to alternatives such as DPF.
\S\ref{sec:offline-basic-composition} gives examples of inefficient DPF operation with multiple blocks and alpha orders.
However, DPF does not {\em always} behave inefficiently when invoked on multiple blocks or with multiple alpha orders.
For example, if all the tasks in \F\ref{fig:dpf_multi_block_killer} uniformly demanded three blocks, then DPF would make the optimal choice.
The same would happen if all the tasks in \F\ref{fig:example-rdp} had RDP curves that were all ordered in the same way across alphas, so that the ordering of highest demands is the same as the ordering of demands at the best alpha order.
In such cases, \ourmethod's ``intelligence'' -- its appropriate treatment of the multiple blocks and focus on the best alpha -- would not provide any benefit over DPF.

Instead, \ourmethod should be expected to improve on DPF when the workload exhibits {\em heterogeneity} in one or both of the following two dimensions: (1) number of demanded blocks and (2) best alphas.
(1) The example in \F\ref{fig:dpf_multi_block_killer} exhibits high heterogeneity in demanded blocks, with Task~1 demanding three blocks while all the others demanding just one block.
(2) The example in \F\ref{fig:example-rdp} exhibits heterogeneity in the best alpha for the different curves.
In evaluation (\S\ref{sec:eval-offline-microbenchmark}), we demonstrate this effects using a microbenchmark that is able to explore a wide range of more or less heterogeneous workloads, showing that indeed, in workloads with more heterogeneity \ourmethod significantly outperforms DPF while in cases of homogeneity among all dimensions, \ourmethod performs similarly to DPF.

For real-world DP ML workloads, we believe it is likely that heterogeneity of demands in both dimensions -- number of blocks and best alphas -- would be realistic.  For example, a pipeline that computes some summary statistics over a dataset might run daily on just the latest block, while a large neural network may need to retrain on data from the past several blocks.
This would result in heterogeneity in number of demanded blocks.
Similarly, pipelines that compute simple statistics would likely employ a Laplace mechanism, while a neural network training would employ subsampled Gaussian.
This would inevitably result in heterogeneity in best alphas, because, as shown in \F\ref{fig:rdp-curves}, different mechanisms exhibit very different RDP curves.

Thus, \ourmethod is broadly applicable to: (1) systems that exhibit both of these dimensions of heterogeneity (as would DP ML workloads in TFX-like systems, or static SQL databases with multiple partitions); (2) systems that operate on a single block (such as non-partitioned SQL databases) but perform \RDP accounting; (3) systems that operate on multiple blocks but perform other types of DP accounting, including traditional DP.  For all these settings, \ourmethod would provide a benefit when the workload exhibits heterogeneity. 

\section{Implementation}
\label{sec:implementation}

{
We implement \ourmethod in two artifacts that we open-source at \url{https://github.com/columbia/dpack}.
The first is a\\ {\bf Kubernetes-based implementation} of \ourmethod.
}
We extend PrivateKube's extension to Kubernetes in multiple ways.
We add support for batched scheduling (\ie schedule tasks every $T$ time units) and task weights.
We implement \ourmethod, and add support for solving the single block knapsack using Gurobi with the Go goop interface~\cite{goop-gurobi}. %
The Kubernetes-based implementation has 924 lines of Go.

The second artifact is a {\bf simulator} that lets users easily specify and evaluate scheduling algorithms for the offline and online settings under different workloads. 
We use a discrete event simulator \cite{simpy} to efficiently support arbitrarily fine time resolutions.
Users use configuration files to define the workload and resource characteristics to parameterize scheduling for both online and offline cases. For example, they can define block and task arrival frequencies, the scheduling period and the block unlocking rate. The simulator also supports plugging different definitions of efficiency, and different block selection patterns for tasks (policies). Currently, the simulator supports two patterns: a random selection of blocks without replacement, and a selection of most recent blocks.
The simulator has 6,718 lines of Python.

\section{Evaluation}
\label{sec:evaluation}

\begin{table}

    \footnotesize
    \centering
    \begin{tabular}{|c|c|p{0.12\textwidth}|c|c|p{0.05\textwidth}|}
        \hline
                 & {\bf Sec.}                                 & {\bf Workload}  & {\bf Setting} & {\bf Prototype} & {\bf Results}                                         \\
        \hline
        {\bf Q1} & \S\ref{sec:eval-offline-microbenchmark}    & microbenchmark  & offline       & simulator       & \F\ref{fig:heterogeneity_knobs}                       \\
        \hline
        {\bf Q2} & \S\ref{sec:eval-offline-microbenchmark}    & microbenchmark  & offline       & simulator       & \F\ref{fig:offline-microbenchmark}                    \\
        \hline
        {\bf Q3} & \S\ref{sec:eval-online-macrobenchmarks}    & Alibaba, Amazon & online        & simulator       & \F\ref{fig:alibaba_eval}-\ref{fig:amazonreviews_eval} \\
        \hline
        {\bf Q4} & \S\ref{sec:eval-kubernetes-implementation} & Alibaba         & online        & Kubernetes      & \F\ref{fig:alibaba-kubernetes-system}                 \\
        \hline
    \end{tabular}
    \vspace{0.4cm}

    \caption{\small {{\bf Workload and methodology of each evaluation question.}}}
    \vspace{-0.3cm}
    \label{tab:questions-and-experimental-contexts}
\end{table}

We seek to answer four evaluation questions:
\begin{itemize}
    \item[{\bf Q1:}] On what types of workloads does \ourmethod improve over DPF, and how close is \ourmethod to Optimal? %
    \item[{\bf Q2:}] How do the algorithms scale with increasing load? %
    \item[{\bf Q3:}] Does \ourmethod present an efficiency improvement for plausible workloads? How much does it trade fairness?
    \item[{\bf Q4:}] How does our implementation perform in a realistic setting? %
\end{itemize}

These questions are best answered with distinct workloads and settings, summarized in
\T\ref{tab:questions-and-experimental-contexts}. %
First, Q1 and Q2 are best addressed in an offline setting with a simple, tunable workload.
To this end, we develop a microbenchmark consisting of multiple synthetic tasks with distinct RDP curves and a knob that controls the heterogeneity in demanded blocks and RDP curves (\S\ref{sec:eval-offline-microbenchmark}).
Second, Q3 and Q4 require a more realistic, online setting and realistic workloads.
In absence of a production trace of DP ML tasks, we develop a workload generator, called {\em Alibaba-DP}, based on Alibaba's 2022 ML cluster trace~\cite{alibaba-gpu}.
We map the Alibaba trace to a DP ML workload by mapping system metrics to privacy parameters (\S\ref{sec:eval-online-macrobenchmarks}).
While we cannot claim Alibaba-DP is realistic, it is the first {\em objectively-derived} DP task workload generator, and we believe it is a more plausible workload than those previously used in related works.  We plan to release it publicly.
Third, Q1-Q3 are algorithmic-level questions independent of implementation and hence we evaluate them in the simulator.
However, Q4 requires an actual deployment on Kubernetes, so we dedicate the last part of this section to an evaluation on Kubernetes with the Alibaba-DP workload (\S\ref{sec:eval-kubernetes-implementation}).

\subsection{Methodology}
\label{sec:methodology}

\heading{Baselines.} The main baseline, common across all experiments, is {\em DPF}.
We consider two other baselines: {\em Optimal}, which is the exact Gurobi-derived \RK solution for the offline setting, and {\em FCFS} (first-come-first-serve), which schedules tasks in an online setting based on their order of arrival. The former is relevant for offline experiments of small scale (few tasks/blocks), since it is not tractable for larger ones.  The latter is relevant for online experiments only.

\heading{Metrics.}
{\em Global efficiency:} defined as either the number of allocated tasks or the sum of weighted allocated tasks. {\em Scheduler runtime:} measures how fast (in seconds), computationally, a scheduling algorithm is.  {\em Scheduling delay:} measures how long tasks are blocked in the waiting queue, for example because of insufficient unlocked budget or because of the batching period $T$; it is measured in block inter-arrival periods (e.g., if blocks arrive daily, the unit is days).  In real life, the total waiting time for a task will be the scheduling delay plus scheduler runtime; for our experiments, since the two are in different units, we never combine them.  We expect in reality scheduler runtimes to be small compared to scheduling delays, for all the evaluated algorithms except for Optimal.

\heading{Machine.}
We use a server with 2 Intel Xeon CPUs E5-2640 v3 @ 2.60GHz (16 cores) and 110GiB RAM.

\subsection{Offline Microbenchmark (Q1, Q2)}
\label{sec:eval-offline-microbenchmark}

\heading{Microbenchmark.}
We design the microbenchmark to expose knobs that let us systematically explore a spectrum of workloads ranging from less to more heterogeneous in demanded blocks and RDP curve characteristics.
The microbenchmark consists of 620 RDP curves corresponding to five realistic DP mechanisms often incorporated in DP ML workloads: \{{\em Laplace, Subsampled Laplace, Gaussian, Subsampled Gaussian, composition of Laplace and Gaussian}\}.
We sample and parameterize these curves with the following methodology meant to expose two heterogeneity knobs:

{\em Knob }{$\mathit{\sigma_{blocks}}$:} To exercise heterogeneity in requested blocks, we sample the number of requested blocks from a discrete Gaussian with mean $\mu_{blocks}$ and standard deviation $\sigma_{blocks}$.
The requested blocks are then chosen randomly from the available blocks.
Increasing $\sigma_{blocks}$ increases heterogeneity in demanded blocks.

    {\em Knob }{$\mathit{\sigma_{\alpha}}$:} To exercise heterogeneity in best alphas, we first normalize the demands (for a block with initial budget $(\epsilon, \delta) = (10, 10^{-7})$) and enforce that there is at least one curve with best alpha $\alpha$ for each $\alpha \in \{3,4,5,6,8,16,32,64\}$.
Second, we group tasks with identical best alphas to form ``buckets''. For each new task, we pick a best alpha following a truncated discrete Gaussian over the bucket's indexes, centered in the bucket corresponding to $\alpha = 5$ with standard deviation $\sigma_\alpha$.  Third, we sample one task uniformly at random from that bucket.
After dropping some outliers (\eg curves with $\epsilon_{\min} < 0.05$), we rescale the curves to fit any desired value of the average and the standard deviation of $\epsilon_{\min}$ for each best alpha, by shifting the curves up or down.
This scaling lets us change the distribution in best alphas while controlling for the average size of the workload (in a real workload, the value of $\epsilon_{\min}$ might be correlated with best alpha and other parameters).
Increasing $\sigma_\alpha$ increases workload heterogeneity in best alphas.

We explore each heterogeneity knob separately.
First, we vary $\sigma_{blocks}$ while keeping $\sigma_{\alpha}=0$ (\ie all the tasks have best alpha equal to 5) and $\mu_{blocks}=10$.
Second, we vary $\sigma_\alpha$ while keeping $\sigma_{blocks}=0$, $\mu_{blocks}=1$ (\ie all the tasks request the same single block).
In both cases, we keep $\epsilon_{\min}$ constant for all tasks.
We set $\epsilon_{\min} = 0.1$ for the $\sigma_{blocks}$ experiment (to keep the number of tasks small enough to be tractable for \OPT) and $\epsilon_{\min} = 0.005$ for the $\sigma_{\alpha}$ experiment (to have a large number of tasks with high diversity in $\epsilon(\alpha)$).

\begin{figure}[t]
    \centering
    \subfigure[Block heterogeneity]{
        \includegraphics[width=0.44\linewidth]{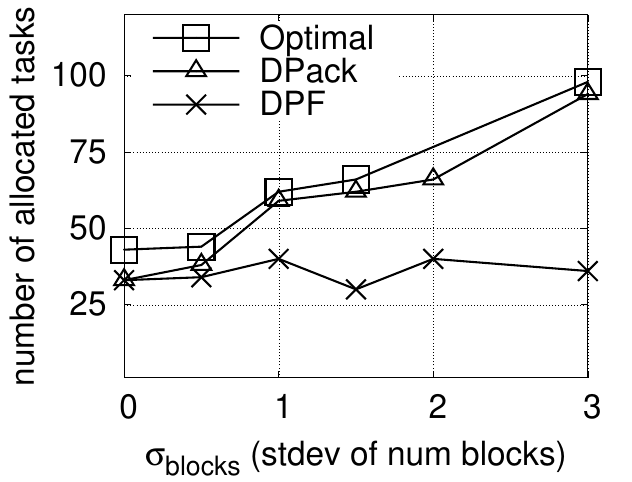}
        \label{fig:blocks_knob}
    }
    \subfigure[Best alpha heterogeneity]{
        \includegraphics[width=0.44\linewidth]{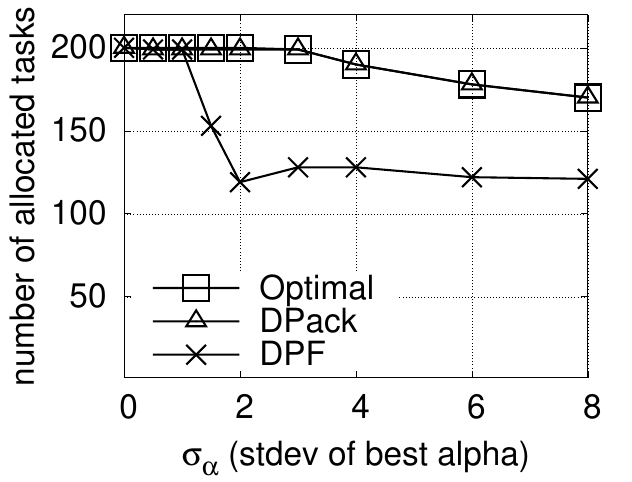}
        \label{fig:alphas_knob}
    }
    \caption{\small {\bf (Q1) \ourmethod under workloads with variable heterogeneity using our microbenchmark.} Shows the global efficiency of the algorithms (y axes) in the offline setting, as heterogeneity increases on the x axes in terms of: (a) variation in number of blocks requested and (b) variation in best alphas for the tasks' \RDP curves.
            {\em Q1~Answer: \ourmethod tracks Optimal closely and significantly outperforms DPF on workloads with high heterogeneity: 0--161\% improvement for \F\ref{fig:blocks_knob} and 0--67\% for \F\ref{fig:alphas_knob}.}
    }
    \label{fig:heterogeneity_knobs}
\end{figure}
\heading{Q1: On what types of workloads does \ourmethod improve over DPF, and how close is \ourmethod to Optimal?}
\F\ref{fig:heterogeneity_knobs} compares the schedulers' global efficiency in the offline setting, as the heterogeneity of the workload increases in the two preceding dimensions: the number of requested blocks (\F\ref{fig:blocks_knob}) and the best alphas of the tasks' \RDP curves (\F\ref{fig:alphas_knob}).
Across the entire spectrum of heterogeneity, \ourmethod closely tracks the optimal solution, staying within 23\% of it.
For workloads with low heterogeneity (up to 0.5 stdev in blocks and 1 stdev in best alphas), there is not much to optimize.
DPF itself therefore performs close to Optimal and hence \ourmethod does not provide significant improvement.
As heterogeneity in either dimension increases, \ourmethod starts to outperform DPF, presenting significant improvement in the number of allocated tasks for over 3 stdev in blocks and 2 stdev in best alphas: 161\% and 67\% improvement, respectively.

As all three schedulers try to schedule as many tasks as they can with a finite privacy budget, these \ourimprovementmicro additional tasks that \ourmethod is able to schedule are tasks that \emph{DPF would never be able to schedule}, because the requested blocks' budget has been depleted for posterity.

\begin{figure}[t]
    \centering
    \subfigure[Scheduler runtime]{
        \includegraphics[width=0.44\linewidth]{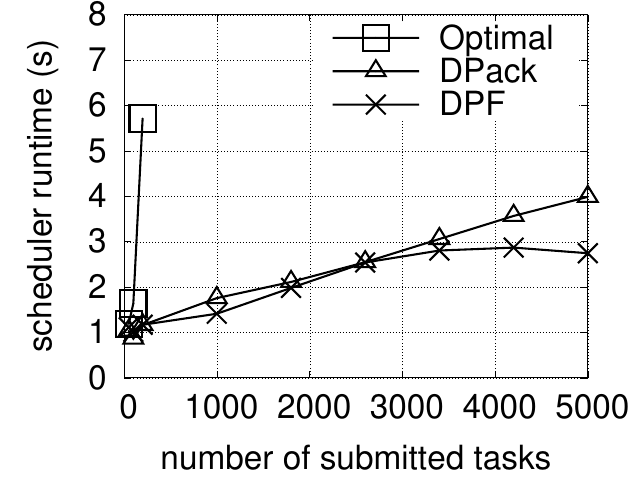}
        \label{fig:offline-microbenchmark-runtime}
    }
    \subfigure[Allocated tasks]{
        \includegraphics[width=0.44\linewidth]{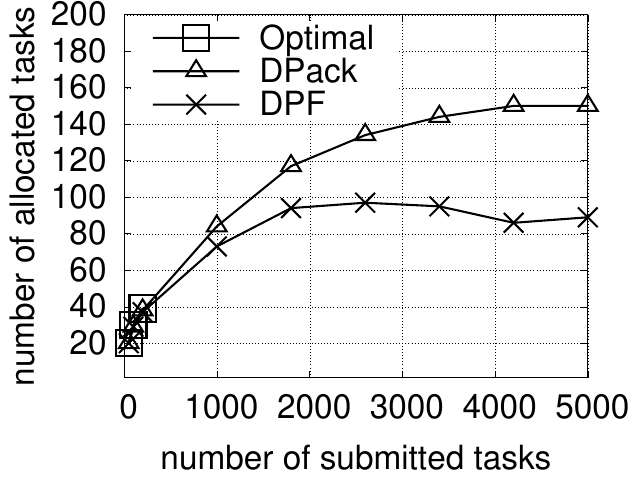}
        \label{fig:offline-microbenchmark-allocated-tasks}
    }
    \caption{\small {\bf (Q2) Scalability under increasing load from the microbenchmark.} (a) Number of allocated tasks and (b) runtime of the scheduler, as function of offered load (x axes).  %
            {\em Q2~Answer: Optimal becomes intractable quickly while \ourmethod and DPF remain practical even at high load.}
    }%
    \label{fig:offline-microbenchmark}
\end{figure}
\heading{Q2: How do the algorithms scale with increasing load?}
\F\ref{fig:offline-microbenchmark-runtime} shows the runtime of our simulator on a single thread. We use a single thread for a fair comparison, but some schedulers can be parallelized (our Kubernetes implementation is indeed parallelized).
We use the microbenchmark with heterogeneity knobs $\sigma_\alpha=4, \sigma_{blocks}=10, \mu_{blocks}=1, \epsilon_{\min} = 0.01$ and 7 available blocks.
Optimal's line stops at $x=200$ tasks because after that its execution never finishes.
\ourmethod takes slightly longer than DPF to run because it needs to solve multiple single-block knapsacks.
\F\ref{fig:offline-microbenchmark-allocated-tasks} shows scheduler efficiency in number of allocated tasks as a function of the number of tasks in the system.
DPF performs the worst, unable to efficiently schedule tasks across multiple blocks and varying alpha order demands.
\ourmethod matches Optimal (up to Optimal's 200 task limit)
and schedules more tasks when it has a larger pool of  tasks to choose from, since it can pick the most efficient tasks.
Since the workload has a finite number of different tasks, as we increase the load, both schedulers reach a plateau where they allocate only one type of task.

\subsection{Online Plausible Workload (Q3)}
\label{sec:eval-online-macrobenchmarks}

We now evaluate online scenarios where tasks and blocks arrive dynamically, and budget is unlocked over time.
The simulator uses a virtual unit of time, where one block arrives each time unit.
Tasks always request the $m$ most recent blocks.
For all the evaluated policies we run a batch scheduler on the available unlocked budget, every $T$ blocks.

\heading{The Alibaba-DP Workload.}
We create a macrobenchmark based on Alibaba's GPU cluster trace~\cite{alibaba-gpu}. The trace includes 1.1 million tasks submitted by 1,300 users over 3 months, and contains each task's resource demands and the resource allocation over time.
We use these metrics as proxies for task DP budget demands, which do not exist in this trace.

We use machine type (CPU/GPU) as a proxy for DP mechanism type.
We assume CPU-based tasks correspond to mechanisms used for statistics, analytics, or lightweight ML (\eg XGBoost or decision trees \cite{dp_decision_tree,dp_xgboost}), while GPU-based tasks correspond to deep learning mechanisms (DP-SGD or DP-FTRL \cite{abadi2016deep, dpftrl}).
We map each CPU-based task to one of the \{{\em Laplace, Gaussian, Subsampled Laplace}\} curves and each GPU-based task to one of the \{{\em composition of Subsampled Gaussians, composition of Gaussians}\} curves, at random.
We use memory usage as a proxy for privacy usage by setting \TDP $\epsilon$ as an affine transformation of memory usage (in GB hours). We don't claim that memory will be correlated with privacy in a realistic DP workload, but that the privacy budget might follow a similar distribution (\eg a power law with many tasks having small requests and a long tail of tasks with large requests).

We compute the number of blocks required by each task as an affine function of the bytes read through the network. Unlike the privacy budget proxy, we expect this proxy to have at least some degree of realism when data is stored remotely: tasks that don't communicate much over the network are probably not using large portions of the dataset.
Finally, all tasks request the most recent blocks that arrived in the system and are assigned a weight of 1. We truncate the workload by sampling one month of the total trace and cutting off tasks that request more than 100 blocks or whose smallest normalized \RDP $\epsilon$ is not in $[0.001, 1]$.
The resulting workload, called {\em Alibaba-DP}, is an objectively derived version of the Alibaba trace.
We use it to evaluate  \ourmethod under a more complex workload than our  synthetic microbenchmark or PrivateKube's also synthetic workload. We open-source Alibaba-DP at \url{https://github.com/columbia/alibaba-dp-workload}.

\begin{figure}[t]
    \centering
    \subfigure[Allocated tasks as function of submitted tasks]{
        \includegraphics[width=0.44\linewidth]{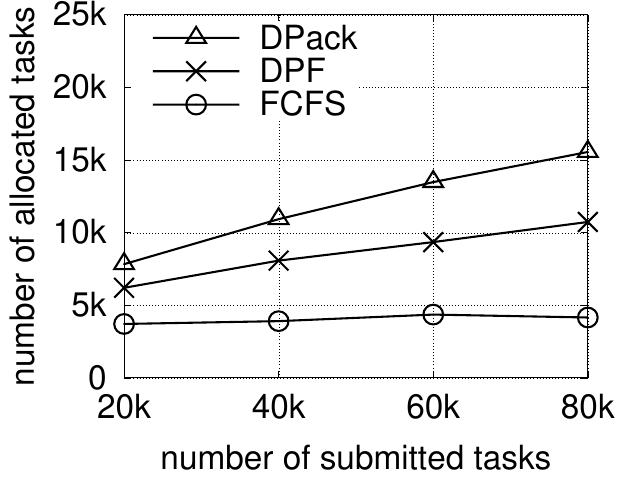}
        \label{fig:alibaba_tasks}
    }~
    \subfigure[Allocated tasks as function of number of available blocks]{
        \includegraphics[width=0.44\linewidth]{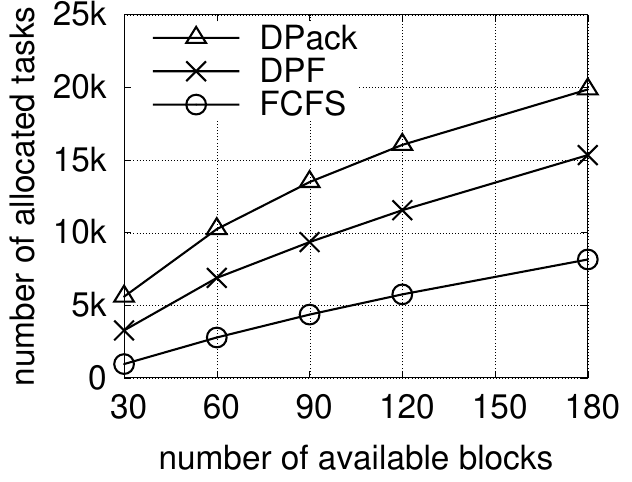}
        \label{fig:alibaba_blocks}
    }~
    \caption{\small {\bf (Q3) Efficiency evaluation on the online Alibaba-DP workload.}
        Number of allocated tasks as a function of (a) offered load for 90 blocks and (b) available blocks for 60k tasks.
            {\em Q3~Answer: Alibaba-DP exhibits sufficient heterogeneity for \ourmethod to present a significant improvement (\ourimprovementalibaba) over DPF.}
    }
    \label{fig:alibaba_eval}
\end{figure}

\heading{Q3: Does \ourmethod present an efficiency improvement for plausible workloads? How much does it trade fairness?}
\F\ref{fig:alibaba_tasks} shows the number of allocated tasks as a function of the number of submitted ones from the Alibaba-DP workload. The results show that as the number of submitted tasks increases, both DPF and \ourmethod can allocate more tasks, because they have a larger pool of low-demand submitted tasks to choose from. This is not the case with FCFS, which does not prioritize low-demand tasks. \ourmethod allocates 22--43\% more tasks than DPF, since it packs the tasks more efficiently.
Similarly, \F\ref{fig:alibaba_blocks} shows the number of allocated tasks as a function of the number of available blocks. As expected, all algorithms can schedule more tasks when they have more available budget. \ourmethod consistently outperforms DPF, scheduling 30--71\% more tasks.
Across all the configurations evaluated in \F\ref{fig:alibaba_tasks} and \ref{fig:alibaba_blocks}, \ourmethod outperforms DPF by \ourimprovementalibaba.
The results confirm that Alibaba-DP, a workload derived objectively from a real trace, exhibits sufficient heterogeneity for \ourmethod to show significant benefit.

The appendix (\F\ref{fig:batch-period-impact}) includes an evaluation of the number of allocated tasks as a function of $T$. We find that for our workloads $T$ has very little impact on the algorithms' performance, and can be set to a low value to minimize scheduling delay.

\heading{Efficiency--Fairness Trade-off.}
While \ourmethod schedules significantly more tasks than DPF on the Alibaba workload, this increased efficiency comes at the cost of fairness, when we use DPF's definition of fairness.
To demonstrate this, we run the Alibaba workload with 90 blocks and 60k tasks, and set the DPF ``fair share'' of tasks to be $\frac{1}{50}$. This means that DPF will always prioritize tasks that request $\frac{1}{50}$ or less of the epsilon-normalized global budget. In the Alibaba trace, using this definition, 41\% of tasks would qualify as demanding less or equal budget than their fair share. With \ourmethod, 60\% of the allocated tasks are fair-share tasks; with DPF 90\% are. However, \ourmethod can allocate 45\% more tasks than DPF.
As expected, this shows that optimizing for efficiency comes at the expense of fairness.
In the case of privacy scheduling, however, due to the finite nature of the privacy budget, DPF's fairness guarantees are limited only to the first $N$ fair share tasks (in the experiment, $N=50$); the guarantees do {\em not} hold for later-arriving tasks.
This makes the overall notion of fairness as defined by DPF somewhat arbitrary and underscores the merit of efficiency-oriented algorithms.

\begin{figure}[t]
    \centering
    \subfigure[Original workload]{
        \includegraphics[width=0.44\linewidth]{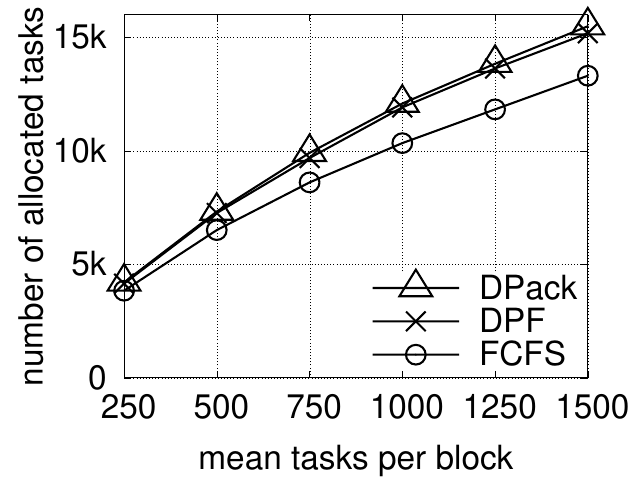}
        \label{fig:amazonreviews_tasks}
    }~
    \subfigure[Workload with task weights]{
        \includegraphics[width=0.44\linewidth]{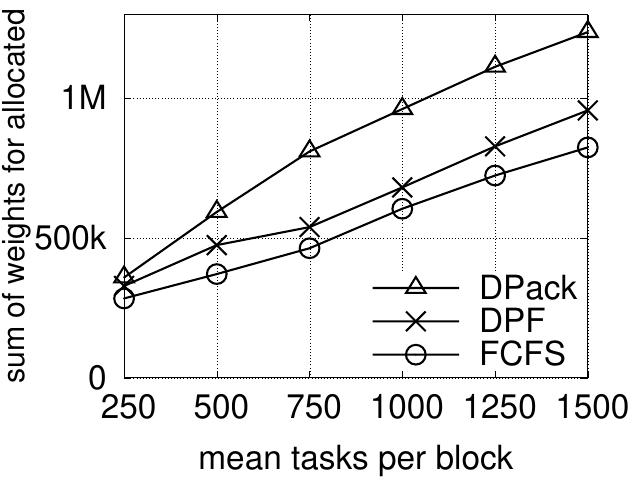}
        \label{fig:amazonreviews_tasks_profits}
    }
    \caption{\small {\bf Evaluation on Amazon Reviews workload from~\cite{privatekube}.}
        (a) This workload, which is synthetic and very simple, exhibits limited heterogeneity, so there is no room for \ourmethod to improve over DPF. (b) Adding randomly selected weights to the tasks creates sufficient heterogeneity for \ourmethod to mark an improvement.  Global efficiency is measured as the sum of weights of allocated tasks (y axis).
    }
    \label{fig:amazonreviews_eval}
\end{figure}

\heading{Another workload: Amazon Reviews~\cite{privatekube}.}
We also evaluate on the macrobenchmark workload from the PrivateKube paper~\cite{privatekube}, which consists of several DP models trained on the Amazon Reviews dataset~\cite{amazon-reviews}.
Unlike our Alibaba-DP, which is rooted in a real ML workload trace, this workload is completely synthetic and very small, and as a result, its characteristics may be very different from real workloads.
Yet, for completeness, we evaluate it here, too.
The workload consists two categories of tasks: 24 tasks to train neural networks with a composition of subsampled Gaussians, and 18 tasks to compute summary statistics with Laplace mechanisms.
Unlike for our Alibaba-DP workload, task arrival needs to be configured for this workload; tasks arrive with a Poisson process and request the latest blocks.
The Amazon Reviews workload has low heterogeneity both in terms of block and the best-alpha variance.
Although tasks request up to 50 blocks, 95\% of the tasks in this workload request 5 blocks or fewer, and 63\% of the tasks request only 1 block.
Moreover, tasks have only 2 possible best alphas (4 or 5), with 81\% of the tasks with a best alpha of 5.
Hence, per our Q1 results in \S\ref{sec:eval-offline-microbenchmark}, we expect DPF to already perform well and leave no room for improvement for \ourmethod.
\F\ref{fig:amazonreviews_tasks} confirms this: all schedulers perform largely the same on this workload.

Next, without modifying the privacy budget or the blocks they request, we configure different weights for submitted tasks, corresponding to different profits the company might get if a task gets to run.
We assume that large tasks (neural networks) are more important than small tasks. Then, we pick an arbitrary grid of weights while still allowing some
small tasks to be more profitable than some large tasks. Weights are chosen uniformly at random from $\{10, 50, 100, 500\}$ for large tasks and $\{1, 5, 10, 50\}$ for small tasks.
This change implicitly scales the number of requested blocks and increases heterogeneity.
In terms of global efficiency, a task with weight $k$ demanding $m$ blocks is roughly similar to a task with weight $1$ demanding $m/k$ blocks. Instead of having most tasks request 1 block, tasks now demand a higher-variance weighted number of blocks (the variation coefficient is 1.9 instead of 1.3).
\F\ref{fig:amazonreviews_tasks_profits} shows the global efficiency, measured as the sum of weights of allocated tasks, as a function of the number of submitted tasks.
Recall that we also incorporate task weights in DPF (\S\ref{sec:offline-basic-composition}). Still, \ourmethod now outperforms DPF by 9--50\%.

\subsection{Kubernetes Implementation Evaluation (Q4)}
\label{sec:eval-kubernetes-implementation}

\begin{figure}[t]
    \centering
    \subfigure[Scheduler runtime]{
        \includegraphics[width=0.44\linewidth]{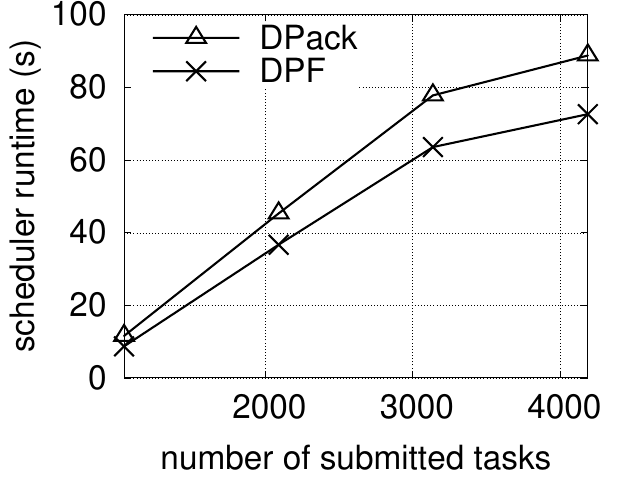}
        \label{fig:alibaba_kubernetes_scheduler_runtime}
    }~
    \subfigure[Scheduling delay CDF]{
        \includegraphics[width=0.44\linewidth]{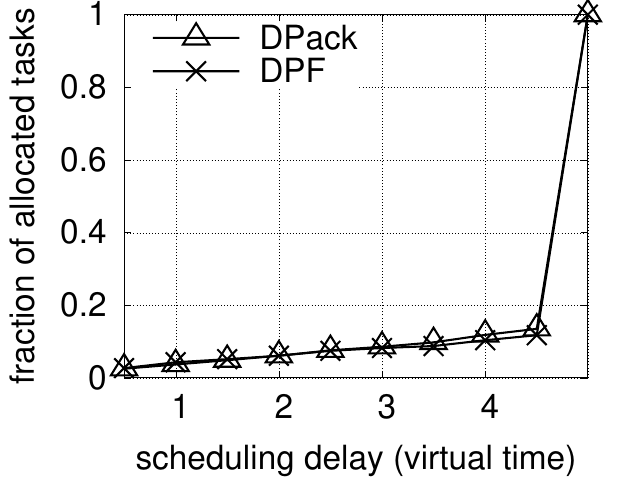}
        \label{fig:alibaba_scheduling_delay}
    }~
    \caption{\small {\bf (Q4) Evaluation on Kubernetes with Alibaba-DP.}
        (a) Scheduler runtime as function of submitted tasks in offline experiment ($T=25$), and
        (b) CDF of scheduling delay (we exclude scheduler runtime) for allocated tasks in online experiment ($T=5$).
            {\em Q4~Answer: \ourmethod has only a modestly higher runtime than DPF, because system-related overheads dominate runtime. In an online setting (b), scheduling delays  are almost identical across schedulers.}
    }
    \label{fig:alibaba-kubernetes-system}
\end{figure}

\begin{table}
    \footnotesize
    \centering
    \begin{tabular}{|l|c|c|c|}
        \hline
        {\bf Scheduler} & {\bf Number of allocated tasks} \\
        \hline
        \ourmethod      & 1269                            \\
        DPF             & 1100                            \\
        \hline
    \end{tabular}
    \vspace{0.4cm}

    \caption{\small {{\bf Efficiency on Kubernetes prototype with Alibaba-DP.} }}
    \label{t:profits-table}
    \vspace{-0.5cm}

\end{table}

\heading{Q4: How does our implementation perform in a realistic setting?}
We evaluate the Alibaba-DP workload on our Kubernetes system.
    {\em Scheduler runtime:} We first estimate the scheduler's overhead by emulating an offline scenario, where all the tasks and blocks are available. To do so, we use a large $T=25$. For this experiment, we generate a total of 4,190 tasks by sampling 2 days of the Alibaba cluster trace. The experiment shows the runtime as a function of the number of submitted tasks. It uses 10 offline and 20 online blocks.
\F\ref{fig:alibaba_kubernetes_scheduler_runtime} shows the total time spent in the scheduling procedure, which includes Kubernetes-related overheads (\eg inter-process communication and synchronization). As noted in \S\ref{fig:offline-microbenchmark-runtime}, \ourmethod has a higher overhead since it solves knapsack subproblems.
\ourmethod has a higher runtime overhead than DPF since it has to recompute the efficiency of each task when the global state changes after a
scheduling cycle, while DPF computes the dominant share of each task only once. Nevertheless, the overhead is modest, because: (a) the Kubernetes overheads dominate, and (b) the \ourmethod (and DPF) algorithms are parallelized. In addition, since DP tasks are often long-running (\eg distributed training of deep neural networks), the scheduling delay of \ourmethod in many cases is insignificant compared to the total task completion time.

    {\em Scheduling delays and efficiency:}
We run an experiment to measure the scheduling delays (\F\ref{fig:alibaba_scheduling_delay}) and efficiency (Table~\ref{t:profits-table}) in an online scenario on Kubernetes.
We use the same workload and number of blocks as in \F\ref{fig:alibaba_kubernetes_scheduler_runtime}, with $T=5$.
As before, \ourmethod is more efficient than DPF.
Scheduling delay, measured in virtual time, excluding scheduler runtime, shows no significant difference between the two policies.

\section{Related Work}
\label{sec:related-work}

We have already covered the details of the most closely related works: DPF and related systems for privacy scheduling~\cite{sage, privatekube, cohere} (background in \S\ref{sec:background:dpf}, efficiency limitations in \S\ref{sec:offline-basic-composition} and \S\ref{sec:offline-rdp-composition}, and experimental evaluation in \S\ref{sec:evaluation}).
To summarize, we adopt the same threat and system models, but instead of focusing on fairness, we focus on efficiency because we believe that the biggest pressure in globally-DP ML systems will ultimately be how to fit as many models as possible under a meaningful privacy guarantee.
The authors of Cohere~\cite{cohere} concurrently developed a privacy management system with novel partitioning and accounting features. They also investigate efficiency-oriented privacy resource allocation, but they rely on an ILP solver for scheduling, which is similar to our {\em Optimal} baseline (\S\ref{sec:methodology}). Their optimal solver faces the same scalability issues we identified, unless tasks query non-overlapping block ranges, thus reducing the number of constraints in the \RK. Cohere supports \ourmethod as an approximate scheduler, and the authors observe that ``the DPK heuristic\footnote{DPack was known as DPK in a previous preprint of our paper.}
achieves within 96\% and 98\% of optimal request volume and utility,
respectively'' on their workload. This further validates \ourmethod.

\heading{Bin packing for data-intensive tasks.}
Multidimensional knapsack and bin packing are classic NP-hard problems~\cite{kou1977multidimensional, azar2013tight, woeginger1997there}.
In recent years, several heuristics for these problems have been proposed to increase resource utilization in big data and ML clusters \cite{tetris, carbyne, graphene, tiresias:nsdi19}.
Some of these heuristics assign a weight to each dimension and reduce to a scalar problem with a dot product \cite{knapsack_problems_textbook, vector_bin_packing}.
We show that the R\'enyi formulation of differential privacy generates a new variation of the multidimensional knapsack problem, making standard approximations and heuristics unsuitable. %

\heading{Scheduling trade-offs.}
Fairness and performance is a classic tradeoff in scheduling even in single-resource scenarios.
Shortest-remaining-time-first (SRTF) is optimal for minimizing the average completion time, but it can be unfair to long-running tasks and cause starvation.
Recent works have shown a similar fairness and efficiency tradeoff in the multi-resource setting \cite{carbyne}.
Although max-min fairness can provide both fairness and efficiency for a single resource, its extension to multi-resource fairness \cite{DRF} can have arbitrarily low efficiency in the worst case \cite{hug}.
In this paper, we highlight the fairness-efficiency tradeoff when allocating privacy blocks among multiple tasks with \RDP.

\heading{Differential privacy.}
The literature on {\em DP algorithms} is extensive, including theory for most popular ML algorithms (\eg SGD~\cite{abadi2016deep,yu2019differentially}, federated learning~\cite{McMahan2018LearningDP}) and statistics (\eg contingency tables~\cite{barak2007privacy}, histograms~\cite{xu2012differentially}), and their open source implementations~\cite{ms-harvard-opendp,idm-diffprivlib,google-dp,tensorflow-privacy,opacus}.
These lower-level algorithms run as tasks in our workloads.
Some algorithms focus on workloads~\cite{hardt2010multiplicative}, including on a data streams~\cite{cummings2018differential}, but they remain limited to linear queries.
Some {\em DP systems} also exist, but most do not handle ML workloads, instead providing DP SQL-like~\cite{Mcsherry:pinq,proserpio2014calibrating,zetasql} and MapReduce interfaces~\cite{roy2010airavat}, or support for summary statistics~\cite{mohan2012gupt}.
Sage~\cite{sage}, PrivateKube~\cite{privatekube} and Cohere~\cite{cohere}, previously discussed, handle ML workloads.

\section{Conclusions}
\label{sec:conclusion}

This paper for the first time explores how data privacy should be scheduled efficiently as a computing resource.
It formulates the scheduling problem as a new type of multidimensional knapsack optimization,
and proposes and evaluates an approximate algorithm, \ourmethod, that is able to schedule significantly more tasks than the state-of-the-art.
By taking the first step of building an efficient scheduler for DP, we believe this work builds a foundation for tackling several important open challenges for managing access to DP in real-world settings, such as supporting tasks with different utility functions, investigating job-level scheduling, and better scheduling of traditional computing resources alongside privacy blocks.

\section*{Acknowledgements}
 This work was supported by NSF grants EEC-2133516, CNS-2106530, CNS-2104292, CNS-2106184, NSERC RGPIN-2022-04469, Google, Microsoft, Sloan Faculty Fellowships, and an Onassis Foundation Scholarship.

\clearpage \newpage
{
    \small
    \bibliographystyle{abbrv}
    \bibliography{bib/abbrev,bib/conferences,bib/refs,bib/news,bib/mosharaf}
}

\clearpage \newpage

\setcounter{page}{1}
\appendix
\makeatletter
\@addtoreset{theorem}{section}
\makeatother

\section{Additional Experiments}

\begin{figure}[ht]
    \centering
    \subfigure[Allocated tasks]{
        \includegraphics[width=0.44\linewidth]{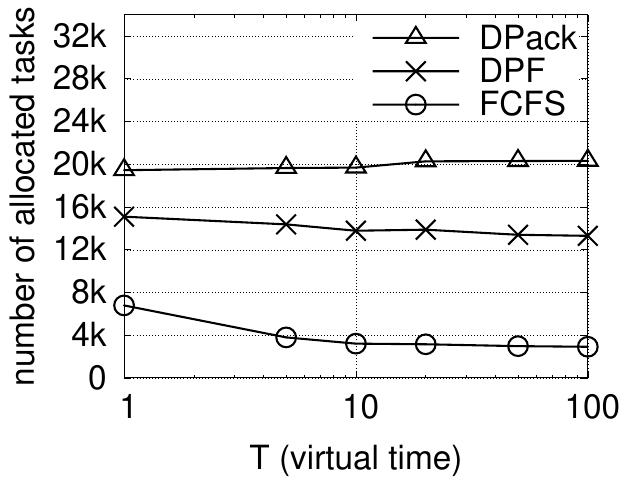}
        \label{fig:batch-period-impact-allocated-tasks}
    }~
    \subfigure[Scheduling delay]{
        \includegraphics[width=0.44\linewidth]{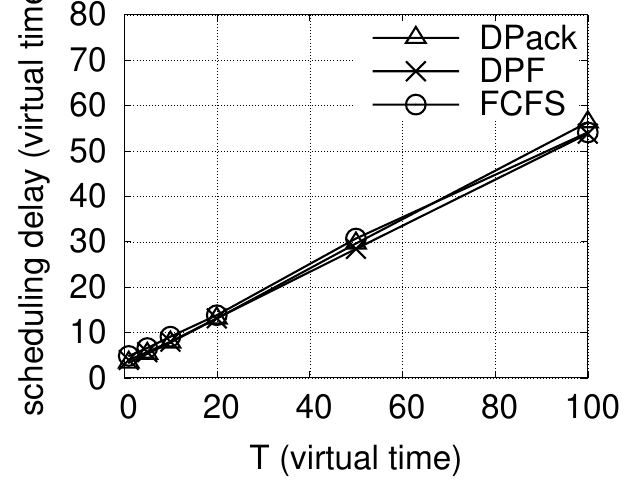}
        \label{fig:batch-period-impact-scheduling-delay}
    }
    \caption{\small {\bf Impact of batching parameter $T$ on global efficiency (a) and scheduling delay (b).}
        (a) Number of allocated tasks, and (b) scheduling delay (in virtual time), both as a function of the batching parameter $T$ (in virtual time).
            {\em In terms of allocated tasks, beyond a reasonable batch size all schedulers are relatively insensitive to the batching parameter.}
        \label{fig:batch-period-impact}
    }
\end{figure}

\F\ref{fig:batch-period-impact}  shows the number of allocated tasks as a function of $T$. With a high $T$, the online setting converges to the offline setting. \ourmethod and DPF perform more or less the same as a function of $T$, while FCFS performs worse. This is because with a high $T$, more budget will be unlocked to schedule large tasks that arrived early, which would otherwise not get scheduled if their budget was not yet unlocked. \ourmethod consistently outperforms DPF by 28--52\% in this experiment.

Therefore, we can conclude that $T$ can be safely set to a relatively low value (to minimize scheduling delay). Indeed, scheduling tasks once per day ($T=1$ in \F\ref{fig:batch-period-impact}, where T is expressed in blocks, corresponding to one day’s worth of tasks) is already giving batches large enough to separate \ourmethod from FCFS.

\section{Formal Proofs}
\label{sec:proofs}
\label{sec:complexity-proofs}

\begin{property}
    The decision problem for the \RK problem is NP-hard.
\end{property}

\begin{proof}
    Consider the decision problem corresponding to \RK: we are given an instance (demands, weights and capacities), and given a sum of weights $t \in \R$ we have to decide whether $t$ is achievable by an allocation.
    We assume the number of \RDP orders $|A|$ is fixed.

    We can build a polynomial time reduction from the knapsack problem (KP) to the \RK problem (PK).

    Consider an instance $(d,w,c)$ of KP with demands $d = (d_1, \dots, d_n)$ and weights $w = (w_1, \dots, w_n)$.
    We define an instance $f(d',w',c')$ of PK with 1 block ($j=1$) and $n$ tasks such that for each task $i \in [n]$ and $\alpha \in A$: $d'_{ij\alpha} = d_i$ (the demands are identical for all the $\alpha$ orders), $c'_{j\alpha} = c$ (idem), and $w'_i = w_i$.

    Thanks to this mapping, we have KP $\le_p$ PK (\ie if we can solve PK, then we can solve KP with polynomial overhead), because $f$ is poly-time computable and $(d,w,c) \in \text{KP} \iff f(d',w',c') \in \text{RK}$. Indeed,
    \begin{itemize}
        \item If we have a KP allocation with sum of allocated weights $t$, we can reuse the exact same allocation in $f(d',w',c')$. The \RK capacity constraint will be satisfied for $\alpha=1$ (for example).
        \item On the other hand, given a PK allocation $f(d',w',c')$ that achieves total weight $t > 0$, we can build an allocation in KP that achieves the same weight by selecting any $\alpha$.
    \end{itemize}
    We know that KP is NP-hard, thus PK is NP-hard.
\end{proof}

\begin{property}
    In the single-block case, there is a fully polynomial time approximation scheme (FPTAS) for the \RK problem.
\end{property}
\begin{proof}
    First, we know that the 1-dimensional Knapsack Problem (KP) is in FPTAS.
    For example, for $1 > \epsilon > 0$ and $n \in \N$, prior work~\cite{knapsack_problems_textbook} gives an $(1-\epsilon)$-approximation algorithm $\mathcal{K}$ for KP with runtime $O(n^3/\epsilon)$ on instances of size $n$.
    Consider a single block $j=1$, $n$ tasks and $|A|$ \RDP orders.
    We can approximate the maximum profit for PK by running $|A|$ instances of $\mathcal{K}$:
    \begin{enumerate}
        \item For order $\alpha$, compute the (approximate) solution $\pmaxhat_\alpha$ for $\max_x \sum_i w_ix_i$ subject to $\sum_i d_{ij\alpha}x_i \le  c_{j\alpha}$ using $\mathcal{K}$.
        \item Then, output the maximum profit $\pmaxhat := \max_{\alpha \in A} \pmaxhat_\alpha$.
    \end{enumerate}
    Since $A$ is fixed, this algorithm also runs in $O(n^3/\epsilon)$ time, and it is an $(1-\epsilon)$-approximation algorithm for PK.
    Indeed, there exists an $\alpha \in A$ such that $\pmax_\alpha \ge \pmax$ where $\pmax := \max_x \sum_i w_ix_i$ subject to $\exists \alpha' \in A, \sum_i d_{ij\alpha'}x_i \le  c_{j\alpha'}$.
    Finally, the output satisfies $\pmaxhat \ge \pmaxhat_\alpha \ge (1-\epsilon) \pmax_\alpha \ge (1-\epsilon) \pmax$.
\end{proof}

\begin{property}
    For $m \ge 2$ blocks, there is no FPTAS for the \RK problem unless P=NP.
\end{property}

\begin{proof}
    We know that there is no FPTAS for the multidimensional knapsack problem (d-KP) when $d \ge 2$, unless P=NP \cite{knapsack_problems_textbook}.
    We can reuse the reduction from Prop.~\ref{prop:single-block-fpta} to solve d-KP from PK with $m=d$ blocks with polynomial overhead.
\end{proof}

\section{Artifact Appendix}

We release an artifact at \url{https://github.com/columbia/dpack}. The artifact contains three main components:
\begin{itemize}
    \item Our fork of PrivateKube (OSDI '21) which implements the DPack scheduler in addition to the original DPF scheduler.
    \item The Python simulator we built to specify and evaluate scheduling algorithms in various settings.
    \item The Alibaba-DP workload, a benchmark to evaluate scheduling algorithms for differential privacy.
\end{itemize}

The repository contains detailed instructions on how to use the simulator with the Alibaba-DP workload.

\end{document}